\documentclass[%
reprint,
superscriptaddress,
amsmath,amssymb,
aip,
longbibliography
]{revtex4-1}
\usepackage[utf8]{inputenc}
\usepackage{graphicx}
\usepackage{epstopdf}
\usepackage{amsmath,amssymb}
\usepackage{xcolor}
\usepackage{graphicx}
\usepackage{dcolumn}
\usepackage{bm}
\usepackage{hyperref}
\usepackage{lipsum,comment}
\usepackage{enumerate}   
\usepackage{array, diagbox}
\usepackage{float}

\begin{document}

\preprint{AIP/123-QED}
\title{
Data-driven analysis of metastability in a stochastic bistable system
}

\author{Ankan Banerjee}
\email{ankan.banerjee@leicester.ac.uk}
\affiliation{School of Computing and Mathematical Sciences, University of Leicester, LE1 7RH, Leicester, UK}
\author{Manuel Santos Guti{\'e}rrez}
\email{m.santosgutierrez@leicester.ac.uk}
\affiliation{School of Computing and Mathematical Sciences, University of Leicester, LE1 7RH, Leicester, UK}
\author{John Moroney}
\email{john.moroney@leicester.ac.uk}
\affiliation{School of Computing and Mathematical Sciences, University of Leicester, LE1 7RH, Leicester, UK}
\author{Valerio Lucarini}
\email{v.lucarini@leicester.ac.uk}
\affiliation{School of Computing and Mathematical Sciences, University of Leicester, LE1 7RH, Leicester, UK}
\affiliation{School of Sciences, Great Bay University, Dongguan, P.R. China}

\date{\today}

\begin{abstract}
We present a methodology to analyze metastable properties of a simple prototypical stochastic bistable system by identifying slow inter-well and fast saddle escape processes using the formalism of the Koopman operator. Instead of studying noise-induced transitions by following the trajectories of the system, we track them by studying the time evolution and the decay rate of the subdominant mode of the Koopman operator, thus in a geometry-agnostic framework. The obtained escape time statistics together with the decay rate are in good agreement with the predictions - both the exponential and subexponential ones -  of large deviation theory in the weak-noise limit, both in equilibrium and nonequilibrium conditions. The subdominant Koopman mode also allows for an accurate reconstruction of the competing basins of attraction. Furthermore, going deeper in the Koopman spectrum, we are able to recognise modes that are associated with intra-well variability as well as with the escape of trajectories from the saddle towards the attractor, both in the equilibrium and nonequilibrium case. Our methodology, being grounded in purely data-driven techniques, could provide a comprehensive, multi-scale framework for studying high-dimensional metastable systems.

\end{abstract}

\maketitle

\newcommand{\norm}[1]{\left\lVert#1\right\rVert}
\newpage

\begin{quotation}
    Noise-induced tipping describes the process of stochastically-driven transitions between metastable states of a complex system. Key aspects of the tipping behaviour are the boundaries that separate the basins of attraction of the metastable states, the mean exit time from a given metastable state, and pathways of the transitions.  While these properties  can be studied using rigorous methods that rely on the knowledge of the underlying dynamical system, a drawback persists for real-world complex systems where only partial information is available in terms of time series of observables. Here, we present a method to study noise-induced tipping in a purely data-driven manner that is based on utilizing the framework of the Koopman operator. As a minimal metastable model, we examine a two-dimensional stochastic bistable system that can feature both equilibrium and nonequilibrium conditions, depending on the choice of a parameter. By taking advantage of Koopman analysis, we identify modes responsible for slow inter-well mixing to fast saddle escape. Such a decomposition yields a direct estimation of the mean exit time and of the boundaries separating the competing basins of attraction, and emphasizes the role of the edge state as a mediator of the noise-induced transitions.
\end{quotation}
\section{Introduction}
Metastability is ubiquitous in complex systems and has attracted great interest in the mathematical literature \cite{Freidlin_book:1998,Bovier2015}. Metastable systems are characterised by the co-existence of multiple competing states for a given value of the system's parameter. Manifestations of metastability include the coexistence of Earth's present-day warm climate state with the alternative ice-covered snowball state~\cite{Pierrehumbert2011,voigt_ClimDyn:2010,lucarini_nonlinearity:2017,gottwald_ClimDyn_56:2021,boers_EnvResLett:2022}, the potential for reversals of the large scale ocean circulation~\cite{Rahmstorf1995,Dijkstra2003,buckley_RevGeoPhys:2016,Boers2021,ditlevsen_NatComm:2023,lohmann_PNAS:2024}, the reversals of solar and Earth's magnetic fields~\cite{berhanu_EPL:2007,tobias_JFM:2021,alberti_GRL:2023} and, generally,  ecological, biological and engineering systems~\cite{serdukova_SciRep:2017,Boers2017,hudetz_FrontiersSysNeuroSc:2014,newby_JMathBio:2014,sujith_Book:2021,Rossi2025,Hastings2010,Morozov2024,Hastings2026}. The metastable features of a dynamical system can often be understood by looking at an energy-like function defined over the phase space, whereby its local minima indicate the presence of metastable states, and its saddle correspond to the edge states embedded in the basin boundaries \cite{lucarini_nonlinearity:2017}. Transitions between metastable states can be realized by stochastic forcing applied to an otherwise multi-stable deterministic system, leading to noise-induced tipping \cite{ashwin_2012}. Rare event algorithms can be used to dramatically improve the statistics of noise-induced transitions ~\cite{bouchet_PRL:2019,Cini2024,Cini2024}.

Often the transitions between metastable states can have catastrophic consequences as one departs from a reference (and usually desired) state and, therefore, it is of the utmost interest to quantify the stability of a given state and its proximity to tipping. A quantitative understanding of the stability of metastable states and their proximity to tipping can be obtained by studying mean exit times~\cite{arani_science:2021,lucarini_PRL:2019} and most probable escape paths~\cite{Giorgini2020,Soons2024,lohmann_PRSA:2025}. 

In the weak-noise limit, large deviation theory~\cite{ellis_PhysicaD:1999} gives a leading order approximation of the mean exit time that grows exponentially with the inverse of the noise strength, apart from a subexponential prefactor. Additionally, transitions occur prevalently along special paths defined as instantons \cite{Freidlin_book:1998,touchette_arXiv:2011,galfi_LaRivista:2021}. Instantons  connect the relevant attractor with a nearby saddle that acts as a gateway for noise-induced transitions \cite{lucarini_PRL:2019,margazoglou_ProcRSA:2021}. Note that escape paths can depart substantially from the instantonic trajectory even in the weak noise limit  in the case of multiscale systems \cite{borner_PRR:2024}. 

Large deviation results on metastable system can mostly be derived via a variational approach, whereby one minimizes an action functional, leading, in particular, to the definition of the quasipotential of the system, which plays the role of the rate function  \cite{grassberger_JPhysA:1989,freidlin_springer_book:1998}. In the case of reversible stochastic diffusions, after a suitable change of variables, it is possible to express the drift term as (minus) the gradient of such quasipotential, which is then simply referred as potential, as discussed below. There is a significant amount of numerical and analytical work for computing quasipotentials, instantons, and sub-exponential prefactors~\cite{grafke_Chaos:2019,paskal_JScComp:2022,grafke_CommPAM:2024,bouchet_Annales:2016,bouchet_JStatPhys:2022}. Rigorous estimate of the pre-exponential factor has been obtained relatively recently for the reversible case \cite{Helffer2004,BGK2005}, whilst ~\cite{bouchet_Annales:2016,bouchet_JStatPhys:2022} computed pre-exponential corrections for irreversible systems which depend on the non-Gibbsianness of the system along the instanton. 

Paskal and Cameron~\cite{paskal_JScComp:2022} developed an algorithm, based on higher order interpolation of the quasipotential and its gradient between mesh points on the two-dimensional phase space. Despite achieving higher order accuracy, these methods are often cursed by the dimensionality of the system and require the knowledge of the drift term of the system for the estimation of statistical quantities, mean exit times, transition times and quasipotentials. Instead, recent progress has been made to estimate mean exit times and escape probabilities based on deep learning methods~\cite{li_springer:2024,liu_Chaos:2025}.

It seems relevant to address whether it is possible to augment the large deviation theory-based framework discussed above with mathematical techniques that are more amenable to treatment in a data-driven way. 

In the last two decades, Koopman theory has emerged as a rather powerful analytical approach to the study of complex systems. The fundamental idea is to move from the study of the trajectories to the study of suitably defined observables of the system
\cite{mezic_PhysicaD:2004,mezic_Nonlin_Dyn_41:2005,Budisic2012,Brunton_SIAMreview:2021}. Using the framework of the Koopman operator one can derive, roughly speaking, an equivalence between the nonlinear evolution of the state variables and a linear evolution in the (infinite) space of observables.

The numerical approximation of Koopman operators relies on finding a suitable correlation between observables and their image under a single-step time advancement. Dynamic mode decomposition (DMD)~\cite{schmid_DMD:2008} and its various advancements~\cite{schmid_ARFM:2022,Budisic2012,Brunton_SIAMreview:2021,colbrook_chapter:2024} have emerged as reliable algorithms to approximate Koopman operators. Rowley et al.~\cite{rowley_JFM:2009} showed a relationship between Koopman mode decomposition and DMD, specifically that DMD modes constitutes a subset of Koopman modes. Later on, Williams \textit{et al.} developed extended DMD (EDMD)~\cite{williams_JNSc:2015} and kernel DMD~\citep{williams_IFAC:2016} which extends the applicability of DMD methods for a wide range of problems. Recent works by Colbrook and collaborators~\cite{colbrook_CommPAM:2024,colbrook_JFM:2023} introduce residual DMD (ResDMD) algorithms that has augmented the accuracy of finite-dimensional approximations of Koopman operators by computing the residual (associated with the full infinite-dimensional Koopman operators) of the approximated Koopman operator itself which greatly reduces the generation of spurious spectra.

This operator theoretic framework has been used for system identification ~\cite{Brunton2016,Lusch2018}, to identify coherent structures in oceans~\cite{froyland_PRL:2007,Chen_NatClimAtmosSc:2024}, to differentiate causality and coherence in turbulent channel flow~\cite{jimenez_JFM:2023}, to develop machine-learning-based surrogate models to analyze the dynamics of turbulent flows in a Rayleigh-Benard Convection system~\cite{Markmann_IEEE:2024}, to discover and characterize causal signals~\cite{Nathaniel_commphys:2025}, to derive coarse-graining methods for multiscale systems \cite{santos_Chaos:2021}, to design controllers for complex systems~\cite{korda_Springer:2020,ottoARcontrol:2021}, to name a few applications. Recently, Koopman theory has also been used to derive an interpretable version of the fluctuation-dissipation theorem, to obtain efficient response operators, and to explain the occurrence of criticality and tipping behaviour in complex systems \cite{santos2022,lucarini_NatRevPhys:2023,zagli_SIAM:2026,lucarini:arXiv:2025,Lucarini2026CSF}.

Motivated by these works, here we propose to  investigate metastability by combining analytical and data-analysis methods based on large deviation theory with the analysis of the key dominant modes and eigenvalues of Koopman operator and its adjoint counterpart, the Perron-Frobenius operator \cite{lasota}. Specifically, following the results obtained in \cite{matkowsky_JAM:1981,Bovier2005} for equilibrium systems and by \cite{le_CommMatPhys:2024} for the general case of nonequilibrium systems, we address the problem of linking the spectral properties of data-driven approximations of the Koopman operator with quantitative characterizations of metastability and the temporal statistics of noise-induced tipping on a simple yet instructive model that can be tuned in such a way to represent both equilibrium and non-equilibrium  conditions. 

The rest of the paper is organised as follows: we start with a short description of large deviation principles and introduce the system in section~\ref{sec:ldt_preliminaries} followed by a description of Koopman methodologies in section~\ref{sec:transfer_operators}, then we analyse our findings in the result section~\ref{sec:results}, and we present out conclusions in section~\ref{sec:conclusions}. \hyperref[sec:appendixI]{Appendix A} - \hyperref[sec:appendixII]{Appendix B} provide more detailed information on the spectral methods used in this study. 

\section{Escape Processes in Metastable Systems}
\label{sec:ldt_preliminaries}
 In order to ensure that our text is self-contained and to introduce the notation, we first briefly discuss some key elements of large deviation theory that are central to the study of escape processes in metastable systems. A much more detailed discussion can be found, for instance, ~\cite{touchette_arXiv:2011,berglund_MarkovPros:2013,grafke_Chaos:2019,galfi_LaRivista:2021}.
We consider an It\^{o} elliptic diffusion process of the form:
\begin{equation}
d{\pmb{x}(t)} = {\pmb{F}(\pmb{x}} (t)) dt + \alpha {\pmb{s}(\pmb{x}} (t)) d {\pmb{W}_t},
\label{eq:stochastic_DE}
\end{equation}
where, ${\pmb{x}}(t)\in\mathbb{R}^N$, ${\pmb{F}:\mathbb{R}^N\rightarrow\mathbb{R}^N}$ is a smooth drift vector field, ${\pmb{W}}_t$ is a vector of $N$ independent Wiener processes, and $\pmb{s}:\mathbb{R}^{N\times N}\rightarrow\mathbb{R}^N$ multiplied times the noise amplitude $\alpha>0$ defines the noise law. We impose that the noise is non-degenerate, thus requiring that the $N\times N$ covariance matrix ${Q(\pmb{x})}=\pmb{s}(\pmb{x})\pmb{s}^T({\pmb{x}})$ is invertible $\forall~ \pmb{x}$. 

Probability density functions evolve according to the Fokker-Planck (FP) equation \cite{Risken89,Pavliotis2014}:
\begin{equation}
    \partial_t \rho=\mathcal{L}^\star \rho = \frac{\partial}{\partial x_i} (-F_i \rho +\frac{\alpha^2}{2} \frac{\partial}{\partial x_j}(Q_{ij}  \rho)),
    \label{eq:FPE}
\end{equation}
where the FP operator $\mathcal{L}^*$ generates a strongly continuous semigroup. We assume that  it has a simple zero eigenvalue, whose eigenfunction $\rho_1$ is the density associated with the unique invariant measure of Eq.~\eqref{eq:stochastic_DE}, namely, $\mathcal{L}^\star \rho_1(\pmb{x})=0$. 

From now on, we assume for simplicity that the covariance matrix $Q$ is the identity matrix and assume that the deterministic part of the dynamics Eq.~\eqref{eq:stochastic_DE} possesses multiple stable fixed points whose basins of attraction are separated by smooth basin boundaries where saddles of the system exist. Stochastic perturbations can force a trajectory to escape from its residing basin of attraction. Though the escape is random, in the weak noise limit the escape path and the rate of escape follow almost deterministic rules due to large deviation principles: escape occurs through a most probable path connecting an attractor to an edge state. This path is called an instanton. Furthermore, mean exit times are estimated by a leading order exponential term with a scalar rate function~\cite{touchette_arXiv:2011,bouchet_Annales:2016, lucarini_nonlinearity:2017}, as detailed below.

If $\pmb{F}(\pmb{x})=-\pmb{\nabla} U({\pmb{x}})$, the system obeys reversible dynamics whereby instantonic trajectories obey the evolution equation $\dot{\pmb{x}}=\pmb{\nabla} U({\pmb{x}})$ and geometrically coincide with relaxation trajectories from saddle to attractors, which instead obey the evolution equation $\dot{\pmb{x}}=-\pmb{\nabla} U({\pmb{x}})$. The attractors are located at the minima of $U$ while the saddles correspond to saddles of the scalar function $U$. Let's assume that the system have no degeneracies so  that these sets are isolated points. The probability of a trajectory leaving the neighbourhood of one  attractors  $\bar{\pmb{x}}_1$ and reaching a second attractor $\bar{\pmb{x}}_2$ is connected to the work done by the fluctuations to overcome the lowest confining potential barrier $\Delta U = U(\pmb{x}_*)-U(\bar{\pmb{x}}_1)$, where $\pmb{x}_*$ is a saddle separating $\bar{\pmb{x}}_1$ and $\bar{\pmb{x}}_2$. The approximation of the mean transition time is given by the Kramers' law and corrections thereof~\cite[Eq.~(1.9)]{BGK2005,berglund_MarkovPros:2013}:
\begin{equation}
\mathbb{E} [\tau_{\bar{x}_1 \rightarrow \bar{x}_2}] \sim_{\alpha \downarrow 0} \frac{2\pi}{|\lambda_u|}\sqrt{\frac{|\mathrm{det}(\mathrm{Hess}(U(x_*)))|}{|\mathrm{det}(\mathrm{Hess}(U(x_1)))|}} \exp{\left(\frac{2 \Delta U}{\alpha^2}\right)},
\label{eqn:Kramers_MET}
\end{equation}
where $\tau_{\bar{x}_1 \rightarrow \bar{x}_2}$ is the first hitting time for a stochastic trajectory departing from $\pmb{\bar{x}_1}$ to reach a neighbourhood of $\pmb{\bar{x}_2}$, Hess$(U(x))=\partial^2U(\pmb{x})/\partial x_i\partial x_j$ is the Hessian of $U$ evaluated at $x$ and ``det'' indicates its determinant, and $\lambda_u<0$ is the only negative eigenvalue of Hess($U$) at the saddle $\pmb{x}_*$.

For irreversible systems, the drift vector can be decomposed as a sum of the conservative (or, time-reversible) and the non-conservative (or, time-irreversible) force field as follows:
\begin{equation}
  {\pmb{F}(\pmb{x})=-\pmb{\nabla}} \Phi + \pmb{R},~ \textit{s.t.}~ {\pmb{\nabla}} \Phi \cdot\pmb{R}= 0,
  \label{eq:decomp_drift_vec}
\end{equation}
where $\Phi$ is known as the \textit{quasipotential}. In this case instantonic trajectories from attractors to saddles obey the evolution equation $\dot{\pmb{x}}=\pmb{\nabla} \Phi({\pmb{x}})+\pmb{R}$ and do not geometrically coincide with relaxation trajectories from saddles to attractors, which instead obey the evolution equation $\dot{\pmb{x}}=-\pmb{\nabla} \Phi({\pmb{x}})+\pmb{R}$.

The mean exit time from an attractor is controlled by an exponential law and is determined by the Friedlin-Wentzell (FW) theory~\cite{freidlin_springer_book:1998,grassberger_JPhysA:1989}. Bouchet and Reygner~\cite{bouchet_Annales:2016} provide an expression for subexponential prefactor. In the simple situation where attractors are isolated points, the mean transition time between two attractors $\pmb{\bar{x}_1}$ and $\pmb{\bar{x}_2}$ going through the saddle $\pmb{x}_*$ is approximated as  
\begin{align}
\mathbb{E} [\tau_{\bar{x}_1 \rightarrow \bar{x}_2}] &\sim_{\alpha \downarrow 0} \frac{2\pi}{|\lambda_u|}\sqrt{\frac{|\mathrm{det(Hess(}\Phi(x_*)))|}{\mathrm{det(Hess(}\Phi(\bar{x}_1)))}} \nonumber\\
&\exp\left(\int_{-\infty}^{\infty} G_n(\gamma_t) dt\right) \exp\left(\frac{2\Delta\Phi}{\alpha^2}\right),
\label{eqn:bouchet_reygner_correction}
\end{align}
where $\gamma_t$ is the instanton path, $G_n=\pmb{\nabla}\cdot \bf{R}$, whilst here the unstable eigenvalue of the drift $\mathbf{F}$, linearized at the saddle, is denoted by $\lambda_u$. In the general nonequilibrium case, the quasipotential $\Phi$ essentially takes the place $U$ has in the equilibrium case in determining the time-statistics of noise-induced transitions. Additionally, the function $G_n$ measures the non-Gibbsiannes of the system, i.e. $G_n$ is zero if and only if the invariant density of the system is proportional to the exponential of the quasipotential divided by the noise amplitude~\cite{bouchet_Annales:2016}. 

\subsection{Our Numerical Model: A double-well system}
\label{sec:large_deviation_principles}
We confine our analysis to the bounded domain $\mathcal{M}=[-1.5,1.5]\times[-1.5,1.5] \subset \mathbb{R}^2$. We construct $\pmb{F}$ following Eq.~(\ref{eq:decomp_drift_vec}) by selecting a potential function $\Phi(x,y) = \frac{x^4}{4}-\frac{x^2}{2}+\frac{y^2}{2}$. This quartic function is a classical double well potential (DWP) model arises in many symmetry breaking transitions. Models based on DWPs are significant in studying metastable states in various systems such as chemical kinetics, phase transitions, transport theory, and mechanical oscillators~\cite{thompson_PRS_A_421:1989}. The transverse component of  $\pmb{F}$ is constructed by rotating by $\pi/2$ the gradient component and rescaling its length by the factor $\mu$, so that $\pmb{R}=\mu\hat{k} \wedge (-{\pmb{\nabla}} \Phi)$. Henceforth, we will interchangeably refer to $\pmb{R}$ as the non-conservative force, the transverse component, or the rotational component. The rotational component makes the system a non-gradient and non-equilibrium one if $\mu\neq0$. Inclusion of the rotational component deforms the phase portrait as if someone has twisted the stable manifold of the saddle point in the anti-clockwise (clockwise) direction if $\mu>0$ ($\mu<0$). Expanding component wise, the drift term is
\begin{eqnarray}
       & F_x = x-x^3+\mu y \nonumber\\
    & F_y = \mu(x-x^3)-y.   
    \label{eqn:sys2}
\end{eqnarray}
The Jacobian matrix of the rotating double well potential (RDWP) system(\ref{eqn:sys2}) is given by 
\begin{equation}
    J_{(x,y)} = \begin{pmatrix}
1-3x^2 &  \mu\\
\mu(1-3x^2) & -1
\end{pmatrix}.
\label{eqn:Jacobian_RDWP}
\end{equation}
Fixed points of this system are $(0,0)$ and $(\pm1,0)$. Eigenvalues of the linearized system at the attractors $(\pm1,0)$ are $-\frac{3}{2}\pm\frac{1}{2}\sqrt{1-8\mu^2}$. The eigenvalues at the saddle $(0,0)$ are $\lambda_{s/u}=\pm\sqrt{1+\mu^2}$, where $u$ and $s$ indicate unstable and stable, respectively. Since all eigenvalues have non-zero real parts, the fixed points remain  hyperbolic for all rotation rates. Therefore, stability of attractors is not affected by rotation, however, stable attractors become stable focus already for weak rotation because as $\mu^2>1/8$ as their eigenvalues feature a non-vanishing imaginary component. Instead, stronger rotations lead to the strengthening of both the contraction and expansion near the saddle.

The corresponding It$\hat{\mathrm{o}}$ diffusion process of Eq.(\ref{eqn:sys2}) is written as 
\begin{eqnarray}
    dx &=& (x-x^3+\mu y)dt+ \alpha dW_{t}^{x} \nonumber\\
    dy &=& (\mu(x-x^3)-y)dt+\alpha dW_{t}^{y}.   
    \label{eqn:Ito_diffusion_process}
\end{eqnarray}
where $\pmb{x}(t)=[x,y]^T \in \mathcal{M}$ denotes a random process at time $t \geq0$, $\alpha$ modulates the intensity of the stochastic forcing, $W_{t}^{x}$ and $W_{t}^{y}$ are independent Weiner processes. Notice that, the noise covariance matrix $Q$ becomes the identity matrix of rank 2.

For systems of the form Eq. (\ref{eqn:Ito_diffusion_process}), the mean transition time from one attractor to the other one depends on the mean exit times to reach the saddle, on the average time spent in the neighbourhood of the saddle, and on the probability $p$ of crossing over to the other attracting basin once the trajectory is at the saddle. Given the symmetry of the system considered here, $p=1/2$ and it takes on the average two attempts before managing to escape, once a trajectory reaches the neighbourhood of the saddle. Therefore, the mean transition time can be estimated as
\begin{equation}
    \mathbb{E} [\tau_{\bar{x}_{1} \rightarrow \bar{x}_{2}}]  \sim 2\mathbb{E} [\tau_{\bar{x}_{1,2} \rightarrow \bar{x}_*}] + \tau_{x_*},
    \label{eqn:modified_transition_time}
\end{equation}
where $\tau_{x_*}$ is the time spent near the saddle. Transition state theory indicates that escape from the saddle can be studied by linearizing the equations at the saddle and solving the backward Kolmogorov equation to find the mean first passage time of trajectories escaping towards either attractor. One finds
\begin{align}
\tau_{x_*}&=\frac{\pi}{\sqrt{-\lambda_u \lambda_s-\alpha^2/4}}=\frac{\pi}{\sqrt{1+\mu^2-\alpha^2/4}}\nonumber \\&=1/2\tau_{x_*\rightarrow \bar{x}_1}=1/2\tau_{x_*\rightarrow\bar{x}_2}\label{eqsaddle}.
\end{align}
In the weak noise regime where $\alpha\ll1$, the noise dependence is negligible, indicating that the escape from the saddle is eminently a deterministic and spontaneous process. Here $\tau_{x_*\rightarrow\bar{x}_j}$ $j=1,2$ are the average escape times from the saddle $\pmb{x}_*$ to the neighbourhood of the two fixed points $\bar{\pmb{x}}_1$ and $\bar{\pmb{x}}_2$, and where we have used the symmetry of the system.

Note that in the derivation of Eq.~\eqref{eqn:bouchet_reygner_correction} (as well as in Eq. \ref{eqn:Kramers_MET}) the time spent on the saddle is neglected because of the weak noise approximation. In the present case, $G_n=0$ because $\nabla \cdot \pmb{R}=0$. Therefore, combining Eq.~\eqref{eqn:bouchet_reygner_correction} and Eq.~\eqref{eqn:modified_transition_time} we obtain for the large deviation scaling
\begin{equation}
\mathbb{E} [\tau_{\bar{x}_{1,2} \rightarrow \bar{x}_*}] \sim_{\alpha \downarrow 0} \frac{\pi}{|\lambda_u|}\sqrt{\frac{|\mathrm{det Hess}\Phi(x_*)|}{\mathrm{det Hess}\Phi(\bar{x_1})}} \exp\left(\frac{2\Delta\Phi}{\alpha^2}\right).
\label{eqn:ER_formula}
\end{equation} 
 Substituting the values of determinants of Hessian matrices at the saddle and at one of the attractors, of the modulus of $\lambda_s$, and the height of the potential $(\Delta \Phi =0.25)$, we obtain 
\begin{eqnarray}
\mathbb{E} [\tau_{\bar{x}_{1,2} \rightarrow \bar{x}_*}](\mu) &\sim& \frac{\pi}{\sqrt{2(1+\mu^2)}} \exp\left(\frac{1}{2\alpha^2}\right)\nonumber,\\
&\sim&  {Z}_\mu \exp\left(\frac{1}{2\alpha^2}\right),
\label{eqn:ER_form_for_DWP}
\end{eqnarray}
From now on, we will use $\mathbb{E} [\tau_{\bar{x}_{1,2} \rightarrow \bar{x}_*}](\mu)$ and $\langle \tau_\alpha \rangle(\mu)$ interchangeably.

Two separate effects are captured by Eq. \eqref{eqn:ER_form_for_DWP}. Clearly, transitions between the two attractors of the double-well potential become more frequent as the noise becomes stronger as a result of the large deviation law. As an effect of the rotational component of the drift term of the Langevin Eq.~(\ref{eqn:Ito_diffusion_process}), the basin boundary between two attractors is twisted anti-clockwise (clockwise) for positive (negative) rotation rates. As a result, the two basins of attraction become geometrically more entangled as the rotation increases and transitions between competing metastable states become more frequent. 

\section{Koopman Operator and Metastability}
\label{sec:transfer_operators}
We next introduce the analytical framework used in this work, that will be used also to define the strategy of data-driven analysis described below. It is possible to define a Koopman operator framework for random dynamical systems with Gaussian white noise~\cite{mezic_Nonlin_Dyn_41:2005}, where the evolution of the random variables is a Markov process. In the present context, the stochastic Koopman operator (also known as the Markov semigroup operator, or, the backward Kolmogorov operator), for a system of the form Eq.\eqref{eq:stochastic_DE} is defined as:
\begin{equation}
[\mathcal{K}^t g] ({\pmb{x}}_0) = \mathbb{E}[g({\pmb{x}}(t))|{\pmb{x}}_0],
\label{eq:stochastic_koopman_operator}
\end{equation}
where ${\pmb{x}}_0$ is the initial configuration, $g : \mathcal{M} \rightarrow \mathbb{C}$ represents a complex-valued observable function, and $\mathbb{E}$ denotes expected value of random processes. The infinitesimal generator of the associated stochastic Koopman operator is given as $\mathcal{L}g = \lim_{t \rightarrow 0+}\frac{\mathcal{K}^tg-g}{t} = \frac{dg}{dt} = \sum_{i=1}^{N}\frac{\partial g}{\partial x_i}F_i+\frac{\alpha^2}{2}\sum_{i=1}^{N}\sum_{j=1}^{N}\frac{\partial^2g}{\partial x_i \partial x_j}Q_{ij}$ where $\pmb{F}$ is the drift vector and $Q$ is the noise covariance matrix. In other words, the family of Koopman operators $\{ \mathcal{K}^t \}_{t \in \mathbb{R}^+}$ is represented as $\mathcal{K}^t=\exp(\mathcal{L}t)$. The FP operator $\mathcal{L}^\star$ given in Eq.~(\ref{eq:FPE}) is the adjoint of the generator $\mathcal{L}$ and is the generator of the Fokker-Planck semigroup $\{\mathcal{P}^t\}_{t \in \mathbb{R}^+}$. 

 In what follows, assume that $\varphi_j,~\varphi^\star_j,~j=1,2,...$ are eigenfunctions of $\mathcal{L}$ and $\mathcal{L^\star}$ respectively, corresponding to eigenvalues $\lambda_j$'s, \textit{i.e.} $\mathcal{L} \varphi_j = \lambda_j \varphi_j$, and  $\mathcal{L}^\star \varphi^\star_j = \bar{\lambda}_j \varphi^\star_j$ following the adjoint relation. It is important to note here that, $\mathcal{K}^t$ and $\mathcal{L}$ share the same set of eigenfunctions, whereas eigenvalues of $\mathcal{K}^t$ are $\exp(\lambda_jt)$. Often, we address Koopman eigenfunctions as Koopman modes. For reversible systems the operator $\mathcal{L}$ is self-adjoint in the $L^2$ space weighted by the invariant measure $\rho_1(\pmb{x})$, which implies that all the eigenvalues of  $\mathcal{L}$ (and of  $\mathcal{L}^*$) are real. Otherwise, eigenvalues can be complex. Since complex eigenvalues occur in conjugate pairs, the adjoint operators $\mathcal{L}$ and $\mathcal{L}^*$ have the same set of eigenvalues, $\lambda_j$.

 Under the assumption of the ergodicity of the system, the real part of each $\lambda_j$ should be non-positive and can be ordered in such a way that they monotonically decreasing with index starting from 0, \textit{i.e.} $\Re(\lambda_j) \leq 0 \text{ with } \lambda_1 = 0$ and $\Re(\lambda_j) \geq \Re(\lambda_{j+1}),~ j =1,2,...$. Eigenfunctions corresponding to $\lambda_1$ represent invariants of the dynamics and are called dominant modes. The dominant mode of the Koopman operator is constant. Whereas, the dominant mode of the FP operator $\varphi_1^*$ returns the stationary density function $\rho_1$ of the system Eq.\eqref{eqn:Ito_diffusion_process} as it satisfies the relation $\mathcal{L}^*\varphi_1^* = 0$. Non-zero eigenvalues give the rate at which eigenfunctions of $\mathcal{L}$ and $\mathcal{L}^\star$ approach the invariants of the system. So, these eigenvalues describe the rate of decay of transients of the system. Eigenvalues deeper into the spectrum and far from 0 represent faster decay rates, and govern the dynamics at fast time scales. The leading sub-dominant modes have the slowest decay rates, as the real parts of the corresponding eigenfunctions are closest to zero. 
 
 Physically, the leading sub-dominant modes describe global properties of the system such as rare transitions between metastable states and slow mixing of density between finite partitions of the phase space~\cite{froyland_PRL:2007}. Froyland~\cite{froyland_NonlinDyn:2005} showed that almost-invariant partitions of an attractor can be identified from leading sub-dominant eigenmodes of the Fokker-Planck operator. Eigenvalues for these modes cluster around 0 and their number depends on the number of partitions. Later, in a different work Froyland \textit{et al.}~\cite{froyland_SIAM:2014} showed that, for a bistable system, the subdominant Koopman mode associated with the second eigenvalue of the spectrum identifies basins of attractions of two metastable states as almost-invariant sets. These almost-invariant sets appear with levels of opposite sign and the change of the level occur at the geometric basin boundary between the two sets. For stochastic multistable systems, successive sub-dominant Koopman modes can identify metastable regions as almost-invariant sets~\cite{lucarini_arXiv:2025}, and can detect basin boundaries as a 0-level set. 

This observation leads to one of the key innovative aspects of our analysis. In order to identify transitions between metastable states in a data driven way, it is sufficient to look at the first sub-dominant Koopman mode $\varphi_2$. Therefore, we propose to construct a one-dimensional timeseries $\varphi_2(\pmb{x}(t))$, (where $\pmb{x}(t), t \geq 0$ is the position in phase space obtained from solving Eq.~\eqref{eqn:Ito_diffusion_process}) so as to identify transitions and quantify the statistics of exit times. A change of sign in $\varphi_2(\pmb{x}(t))$ would indicate an exit from one of metastable states, without any need to know the geometry of the quasipotential and the position of the attractors and of the saddle.
 
Additionally, the decay rates of the subdominant modes are related to the stability of these metastable states.
Considering our case of a symmetric bistable system, we can take advantage of an important analytical result. As shown earlier in the reversible case \cite{matkowsky_JAM:1981,Bovier2005} and very recently proved in general nonequilibrium context \cite{le_CommMatPhys:2024}, in the weak noise limit the first nonzero eigenvalue of $\mathcal{L}$ or $\mathcal{L}^*$ is equal to minus the inverse of the escape time from either attractor: 
\begin{equation}
    \lambda_2(\mu)=-\frac{1}{\langle \tau_\alpha \rangle(\mu)}.
    \label{eqn:Matkowsky_identity}
\end{equation}

 Hence, we can investigate Eq.~\eqref{eqn:modified_transition_time} using two different data-driven approaches: (i) empirical estimation from statistics of exit times by evaluating system variables along the first subdominant Koopman mode; and (ii) a direct estimation of the mean exit time from the associated eigenvalue.

\section{Data Analysis: Metastability, Transitions, and Koopman Analysis}
\label{sec:results}
\subsection{Numerical Simulations}
\label{sec:simulation_details}
We numerically simulate Eq.~(\ref{eqn:Ito_diffusion_process}) using the Euler-Maruyama numerical scheme~\cite{platen_SDE_book:2010}. We consider different noise intensities $\alpha$ ranging from $0.2$ to $0.3$ for analyzing transition statistics. In order to understand the effect of the rotational component of the drift term on the transition scenario, we consider the cases $\mu=0$ (no rotation), $\mu=1$, $\mu=5$, and $\mu=10$, where the two latter cases correspond to far from equilibrium systems featuring a very strong rotation. We choose as  time step for the simulations $dt=0.001$ for $\mu =0,~1$ and $dt=0.0001$ for $\mu =5,~10$ for $\alpha \geq 0.25$. Instead, for $\alpha <0.25$, we choose $dt=0.01$ for $\mu = 0,~1$ and $dt=0.001$ for $\mu = 5,~10$. For every pair of $(\alpha,\mu)$ we generate an ensemble of one hundred timeseries with random initial conditions equally populated around each attractor. In each case, we then integrate the system until we see one transition.

\subsection{Computing Koopman eigenvalues and eigenfunctions}
 In this work we will follow EDMD methodology to approximate Koopman operators. We recapitulate the EDMD-based framework used here in \hyperref[sec:appendixI]{Appendix I}, for more details see~\cite{williams_JNSc:2015,klus_JCompDyn:2016}. Using EDMD, we first construct the Koopman matrix $K$ from an ensemble of timeseries numerically generated from the system~(\ref{eqn:Ito_diffusion_process}). In this regard, we first sample a large number of snapshot pairs from the ensemble of time series with a sampling period of $f_t$ which is an integer multiple of the simulation time step $dt$, \textit{i.e.} $f_t=ndt$, where $n$ is a positive integer. We choose $n$ in such a way that the number of samples are large enough to support the bimodal stationary density function over the computation domain $\mathcal{M}$, covering both attractors with their basins of attraction and the saddle. In that way, we preserve the invariant measure, \textit{i.e.} the stationary density function, in the computation of covariance matrices which are explained in \hyperref[sec:appendixI]{Appendix A}). While the number of snapshot pairs $(M)$ are large enough to ensure the convergence of the Galerkin approximation of the Koopman operator, the exact number of snapshot pairs varies depending on the values of $\mu$ and $\alpha$. Also, the sampling period $(f_t)$ depends on the choice of $\mu$ and $\alpha$. A detailed discussion on $M$ and $f_t$ is included in \hyperref[sec:appendixI]{Appendix A}. We have used polynomials of the form \{$x^iy^j:0\leq i+j \leq 12$\} as dictionary functions here leading to a Koopman matrix $K$ of order $91$. Next, we compute eigenvalues $(\sigma_j:j=1,2,...,91)$ and eigenvectors $(\xi_j:j=1,2,...,91)$ of $K$. The procedure is repeated for different values of $\mu$ and $\alpha$ in order to explore the effect of rotation and noise intensity of the Koopman eigen-pairs.

\begin{figure}    
    \includegraphics[width=\linewidth, height=!]{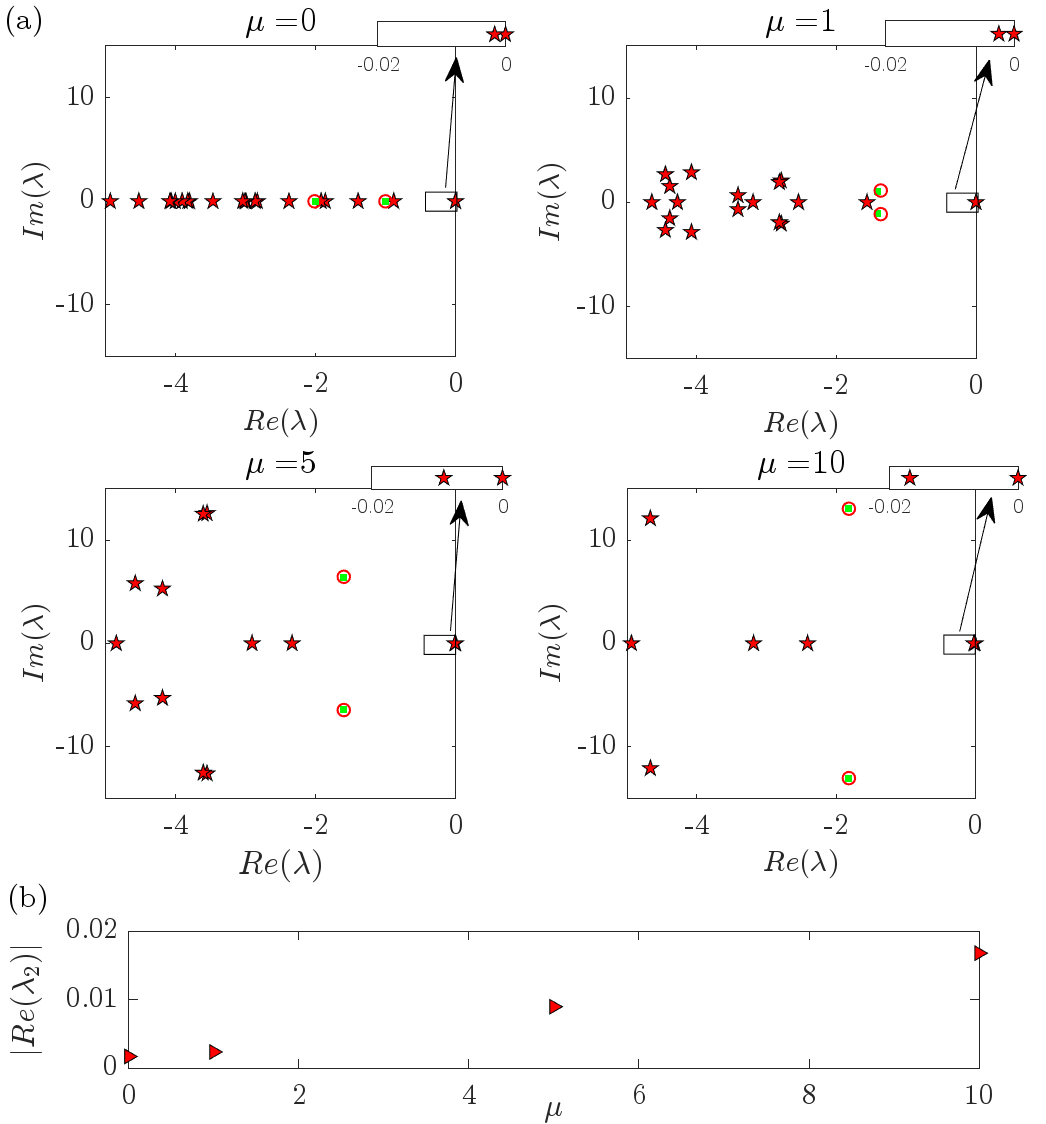}
    \caption{(a) Eigenvalues ($\lambda_j=\log{(\sigma_{j})}/f_t$) of the generator of the semigroup of Koopman operators for different strengths of rotation $(\mu=0,~1,~5,~10)$ corresponding to the noise intensity $\alpha = 0.3$, where $\sigma_j$s are eigenvalues of the approximated Koopman matrix ($K$) and $f_t=0.005$. Solid green squares and hollow red circles are conjugate pairs of eigenvlaues listed in the right column of Table~\ref{tab:coeff_stat_lin_fit_noise0p3}. Leading and subleading eigenvalues are shown in inset of each subplot at the top right corner (b) The magnitude of the real component of the subleading eigenvalue for different values of $\mu$. The plot shows that the spectral gap increases with $\mu$.}
\label{fig:all_mu_KO_eigenvalues_alpha0p4}
\end{figure}

The eigenvalues of $\mathcal{L}$ are $\lambda_j=\log{(\sigma_{j})}/f_t$ and are shown in Fig.~\ref{fig:all_mu_KO_eigenvalues_alpha0p4} for different values of $\mu$, for noise intensity $\alpha=0.3$, and for $f_t=0.005$. These eigenvalues lie on or to the left of the imaginary axis on the complex plane. Consequently, all eigenvalues of $K$ lie inside the unit circle on the complex plane, except the first one, which is real with value 1. Then, eigenfunctions for the Koopman operator $(\varphi_j)$ are approximated from eigenvectors $\xi_j$'s of K. We have also checked with other values of $f_t$ and found that eigenvalue-eigenvector pairs do not differ significantly for different choices of $f_t$. 

The spectral gap is given by the magnitude of the real part of $\lambda_2$ and, in accordance with Eqs. \eqref{eqn:ER_form_for_DWP} and \eqref{eqn:Matkowsky_identity}, increases with the rotation rate $\mu$ and increasing strength of noise. A detailed analysis of $\lambda_2$ is presented further below. Additionally, a faster decay is realised also for the higher-order subdominant modes.

\begin{table}[H]
    \centering
    \begin{tabular}{|c|c|c|}
        \hline
        $\mu$ & Jacobian eigenvalues & Koopman eigenvalues \\
        \hline
        \hline
        0 & $-1, -2$ & $-1.0026, -2.0168$\\
        \hline
        1 & $-1.5 \pm 1.32287i$ & $-1.36273\pm1.13416i$\\
        \hline
        5 & $-1.5\pm7.05337i$ & $-1.59856 \pm 6.45161i$\\
        \hline
        10 & $-1.5 \pm 14.13329i$ & $-1.81455 \pm 13.04829i$\\
        \hline
    \end{tabular}
    \caption{A comparison between eigenvalues of the Jacobian matrix $(-1,0)$, representing the linearized dynamics at one of the stable fixed points, and a pair of eigenvalues of the generator of the Koopman operator. Here, the eigenvalues of $\mathcal{L}$ are computed for  $\alpha=0.3$.}
    \label{tab:coeff_stat_lin_fit_noise0p3}
\end{table}
For $\mu \neq 0$, the first two conjugate pairs of complex eigenvalues, namely, $\lambda_3,~\lambda_4$ and $\lambda_5,~\lambda_6$, approximate eigenvalues of the linearized system $-1.5 \pm \frac{1}{2} \sqrt{1-8\mu^2}$ around the stable fixed points at $(-1,0)$ and $(1,0)$. Note that because of symmetry we have degeneracies in the spectrum. 

Instead, for $\mu=0$, the eigenvalues of the linearized dynamics at both stable fixed points are $-1$ and $-2$. Corresponding eigenvalues of the Koopman spectrum are $\lambda_4,~\lambda_5$ (with  value $\approx-1$), and $\lambda_9,~\lambda_{10}$ (with  value $\approx-2$). These eigenvalues are shown in Fig.~\ref{fig:all_mu_KO_eigenvalues_alpha0p4} as solid green squares and hollow red circles.

The eigenfunctions to these pairs of  (up to multiplicity) eigenvalues (conjugate pairs in case of $\mu \neq 0$) characterize intra-well mixing properties associated with relaxation towards the fixed points. In table~\ref{tab:coeff_stat_lin_fit_noise0p3} we show a comparison between one of the pairs of eigenvalues associated with the fixed point $(-1,0)$ obtained analytically and from Koopman matrices for values of $\mu$  and for noise amplitude $\alpha=0.3$.

\begin{figure}
    \includegraphics[width=\linewidth, height=!]{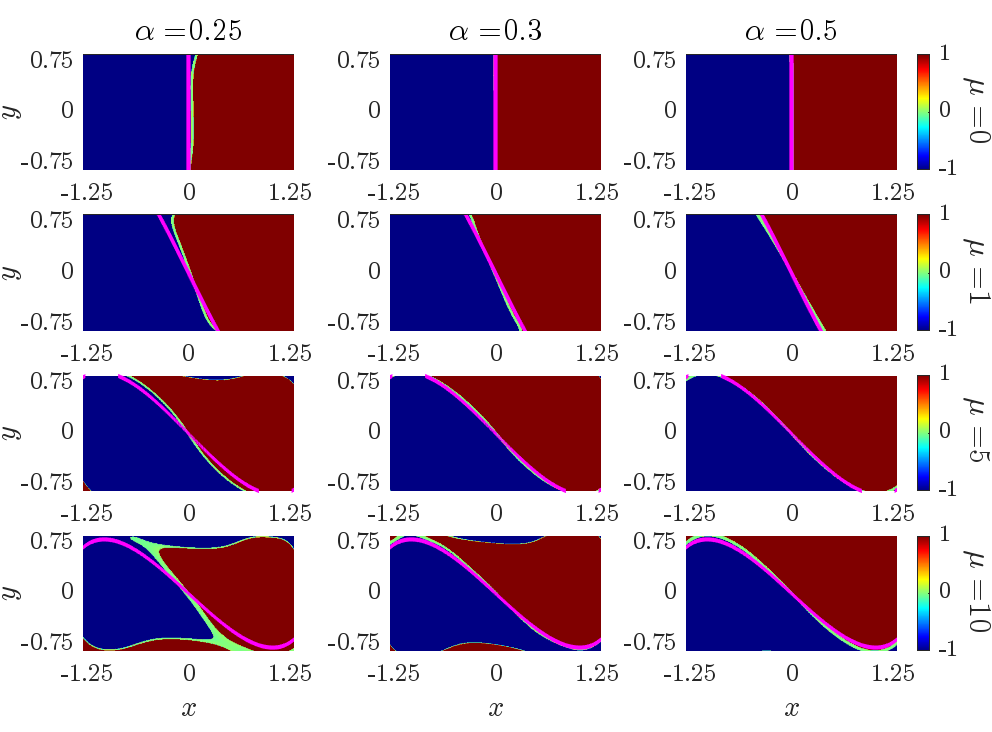}
    \caption{First subdominant mode of the Koopman operator ($\varphi_2$, normalised between -1 and 1) for different values of $\mu$ and $\alpha$.
    } 
    \label{fig:all_x2_comp}
\end{figure}

Following the proposed approach to identify an exit from a metastable state through $\varphi_2$, as discussed in section~\ref{sec:transfer_operators}, we next inspect the structure of $\varphi_2$. We find that $\varphi_2$ partitions the phase space into two almost-invariant sets. Each of these partitions of the phase space has opposite levels in the heat map of $\varphi_2$. Sets with negative (positive) levels contain the metastable states $(-1,0)$ ($(1,0)$) and levels of each set are almost constant. Each of the almost-invariant sets with opposite levels represents a basin of attraction of one of metastable states. For the sake of uniformity, we normalize $\varphi_2$ in such a way that the whole basins of attractions for $(-1,0)$ and $(1,0)$ assumes levels $-1$ and $1$, respectively. In Fig.~\ref{fig:all_x2_comp} we show these partitions for different values of the $(\mu,\alpha)$ pair, as heat maps of $\varphi_2$ on the $xy$-plane, together with the boundary between the basins of attraction of the two attractors, which is depicted by the magenta line.

As can be easily noticed, for each value of $\mu$ there is a very good agreement between the basin boundary and the $0$ isoline (in green) of  $\varphi_2$. Note also that $\varphi_2$ is virtually unaltered as we change $\alpha$, for the fundamental reasons that in all cases it describes the process of relaxation of the system associated with the exchange between the two attractors. See the second row of panels in Fig. \ref{fig:PF_modes_structures} for the corresponding modes of the Fokker-Planck operator. Since $\lambda_2$ describes the slowest decay rate, transitions between blue and red-coloured metastable regions are highly unlikely and each of the red and blue region correspond to quasi-invariant sets. The figure Fig.~\ref{fig:all_x2_comp} also suggests that one can retrieve the boundary between the competing basins of attraction via a purely data-driven approach without having access to the evolution equation of the system, and that this procedure is  possible largely irrespectively of the strength of noise. In other words, one can retrieve partitions of the phase space into disjoint basins of attraction of the underlying deterministic dynamics even from noisy data. 
 
This way of obtaining the basins of attractions of the deterministic counterpart of the system is alternative to studying the 1/2 isosurface of the forward committor function, which could be defined, in our case, as the probability that the process started in a neighbourhood of $(1,0)$ reaches a neighbourhood of $(-1,0)$ before returning to the neighourhood of $(1,0)$ \cite{EVandenEijnden2006,shi_arXiv:2026}. 
 
\subsection{Statistics of the Escape Times}
\begin{figure}[t]
	\centering
	\includegraphics[width=\linewidth, height=!]{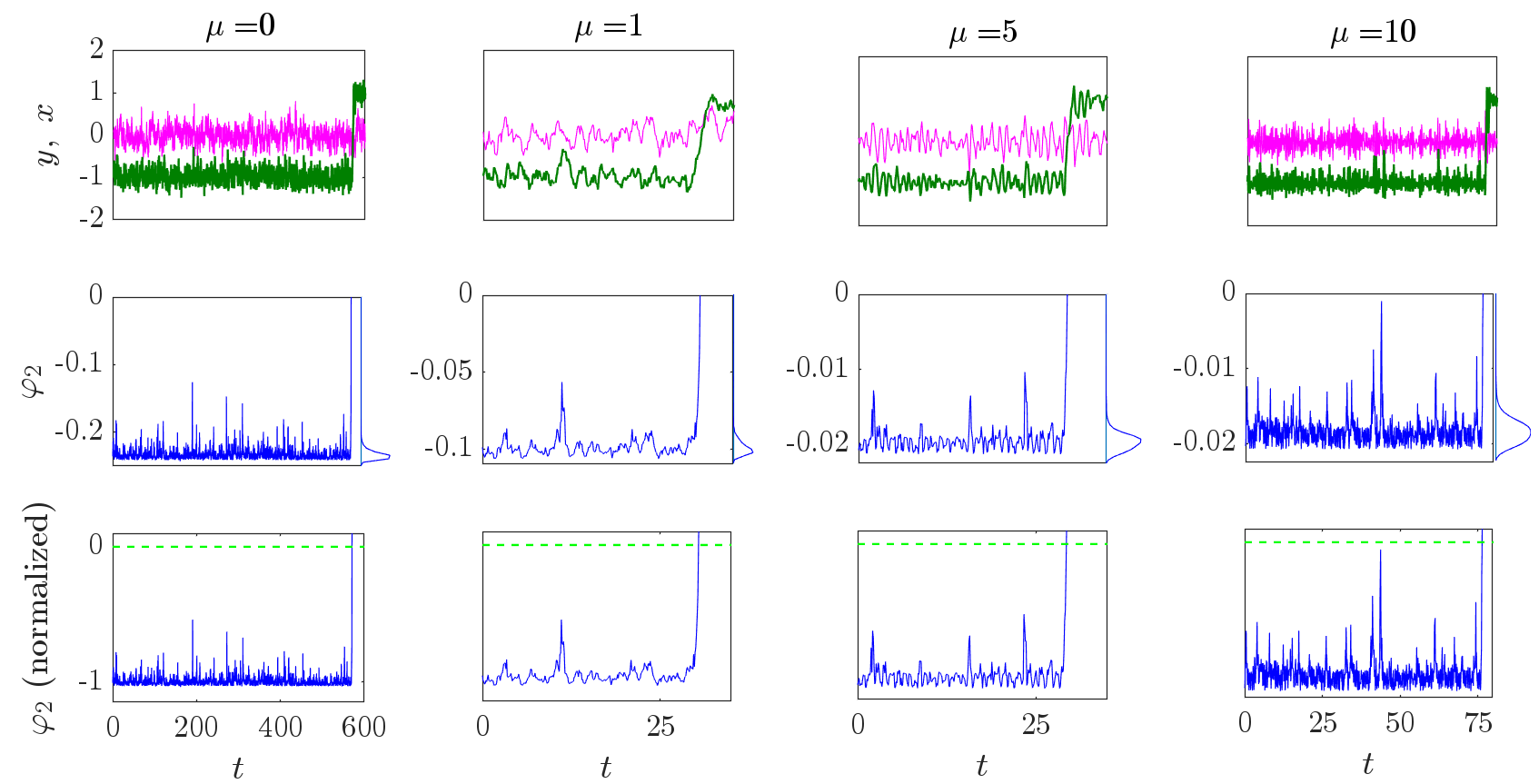}
	\caption{A visual depiction of our method for identifying a transition from one attractor to another. Time series showed here are corresponding to $\mu=0,~1,~5$ and $10$, and $\alpha=0.3$. (Top) - time series of observables; (middle) - the first subdominant Koopman mode $\varphi_2$, evaluated on $x(t)$ , and $y(t)$, density of the time series is also shown on the right of each plots; (bottom) - time series of normalized-$\varphi_2$. Dashed green curves represent basin boundaries at $\varphi_2=0$. A crossing occurs when the normalized $\varphi_2$ crosses the basin boundary.}
	\label{fig:method_of_transition}
\end{figure}
Usually, the numerical investigation of the temporal statistics of noise-induced transitions between metastable states is performed by defining a geometrical neighbourhood around the initial attractor and final attractor and post processing the data in order to assess when a trajectory exits the initial state and reaches the final state. Studying the statistics of escapes from a metastable state is potentially more challenging because one should in principle track the transitions across an extended basin boundary, which is not necessarily well known. In the case one assumes to be in the weak noise limit, one might decide to target the edge state as gateway for the transitions. Nonetheless, this requires knowing where the edge state is situated in phase space. These tasks face increasingly complex challenges as one one faces higher and higher dimensional systems featuring a potentially complex geometry of the boundaries separating the basins of attraction.

\begin{figure}[h]
    \includegraphics[width=1\linewidth]{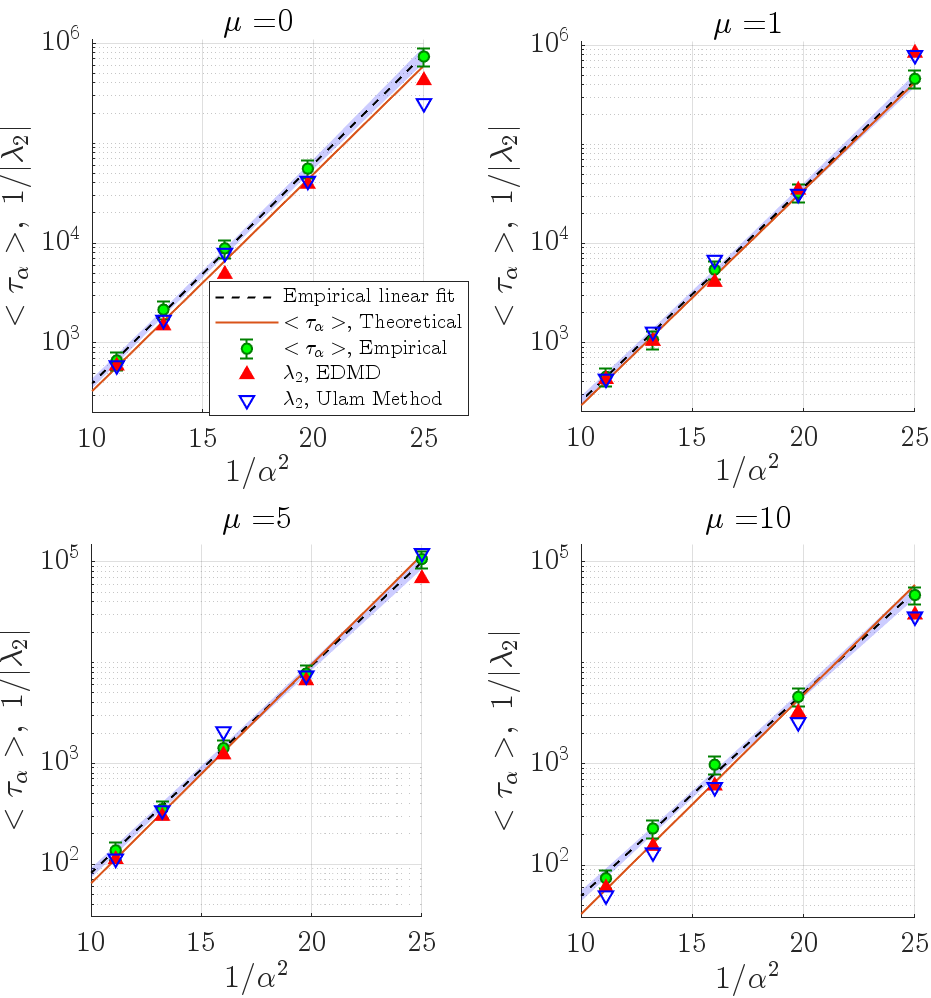} 
    \caption{Test of the linear relationship between the logarithm of $\langle \tau_\alpha\rangle $ and $1/\alpha^2$ for different values of $\mu$. Filled green circles with errorbars (indicating the $95\%$ confidence interval), are the estimation of mean exit times, as computed from ensembles. Dashed black lines are linear fit of these estimators. Shaded regions denote $95\%$ confidence interval of the fit. Solid orange lines are theoretical estimation of exit times - half of mean transition times. Inverse of the modulus of second eigen value of the Koopman operator estimated by EDMD (solid red triangles), inverse of the second eigenvalue of the Markov matrix  computed using the Ulam's method (hollow inverted blue triangles) are also shown here.} 
\label{fig:comp_transition_time_at_different_waiting_times}
\end{figure}

Here we take a different route and look at the escape process by focusing on the value of subdominant Koopman mode as a function of time $\varphi_2(\pmb{x}(t))$. Our approach is particularly simple in this case because the system has a special symmetry but could be easily generalised for more complex geometries and multiple competing metastable states. We then proceed as follows. We initialise the system at the point (-1,0), where $\varphi_2(\pmb{x}(t))\approx-1$. An exit is recorded at $t=t_{esc}$ if $\varphi_2(\pmb{x}(t))$ crosses 0 at $t=t_{esc}$. We then perform a resetting and restart the system very close to the attractor. Given the symmetry of the system, we repeat the procedure by initialising the system almost at the attractor (1,0) and record the escape time when  $\varphi_2(\mathbf{x}(t))$ first crosses zero from positive to negative numbers. A total of 100 escapes (50 for each attractor) is recorded for each choice of parameters. Note that following this strategy we have to look at a simple scalar time series and any geometrical complexity is avoided, as all spatial information is implicitly contained in the level sets of $\varphi_2(\mathbf{x})$, see Fig. \ref{fig:method_of_transition}.

In Fig.~\ref{fig:comp_transition_time_at_different_waiting_times} we show the estimate of the mean exit times $\langle \tau_\alpha \rangle(\mu)$ from the simulation data, where the errorbars centered on the solid green circles indicate the $95\%$ confidence interval, which have been obtained by considering 100 samples for each datapoint. Dashed black lines indicate the best fit of the approximate linear relationship between $\log(\langle \tau_\alpha \rangle)(\mu)$ and $1/\alpha^2$. Uncertainties on the fit have been computed using a bootstrap re-sampling of the mean exit times to generate 1000 sample data points out of ensembles of the transition times computed from numerical simulations. Orange lines show the analytical result given in Eq.~(\ref{eqn:ER_form_for_DWP})). The plot shows that estimated exit times from the data-driven framework are in good agreement with their theoretical counterpart.

Additionally, following the rigorous result presented in \cite{le_CommMatPhys:2024}, we also investigate whether we can approximate the expectation value of the exit time as $1/|\lambda_2|$. Figure ~\ref{fig:comp_transition_time_at_different_waiting_times} shows that we have convincing agreement between $1/|\lambda_2|$ (solid red triangles) and $\langle \tau_\alpha \rangle$. To verify the robustness of the results with respect to the methodology employed for estimating the subdominant eigenvalue of the Koopman operator, we repeat our analysis by discretising the system both spatially and temporally by performing a Ulam partition \cite{ulam1960collection} of the phase space and constructing the stochastic matrix describing the probability of transitions of the trajectory between the various (disjoint) cells of the partitions within a given time interval. The stochastic matrix is an approximation of the Perron-Frobenius operator of the system, and its subdominant eigenvalue provides an approximation of $\lambda_2$. Details on the methodology are given in App. \ref{sec:appendixII}. As discussed in \cite{lucarini_arXiv:2025,Lucarini2026CSF}, the use of the Ulam method boils down to considering as Koopman dictionary the characteristic functions of the various cells that define the partition of the phase space. The results of the analysis performed using the Ulam method are shown as inverted hollow blue triangles in Fig.~\ref{fig:comp_transition_time_at_different_waiting_times}, and are in good agreement with the estimates obtained using the dictionary of monomials.

\begin{figure}[t]
    \includegraphics[width=1\linewidth]{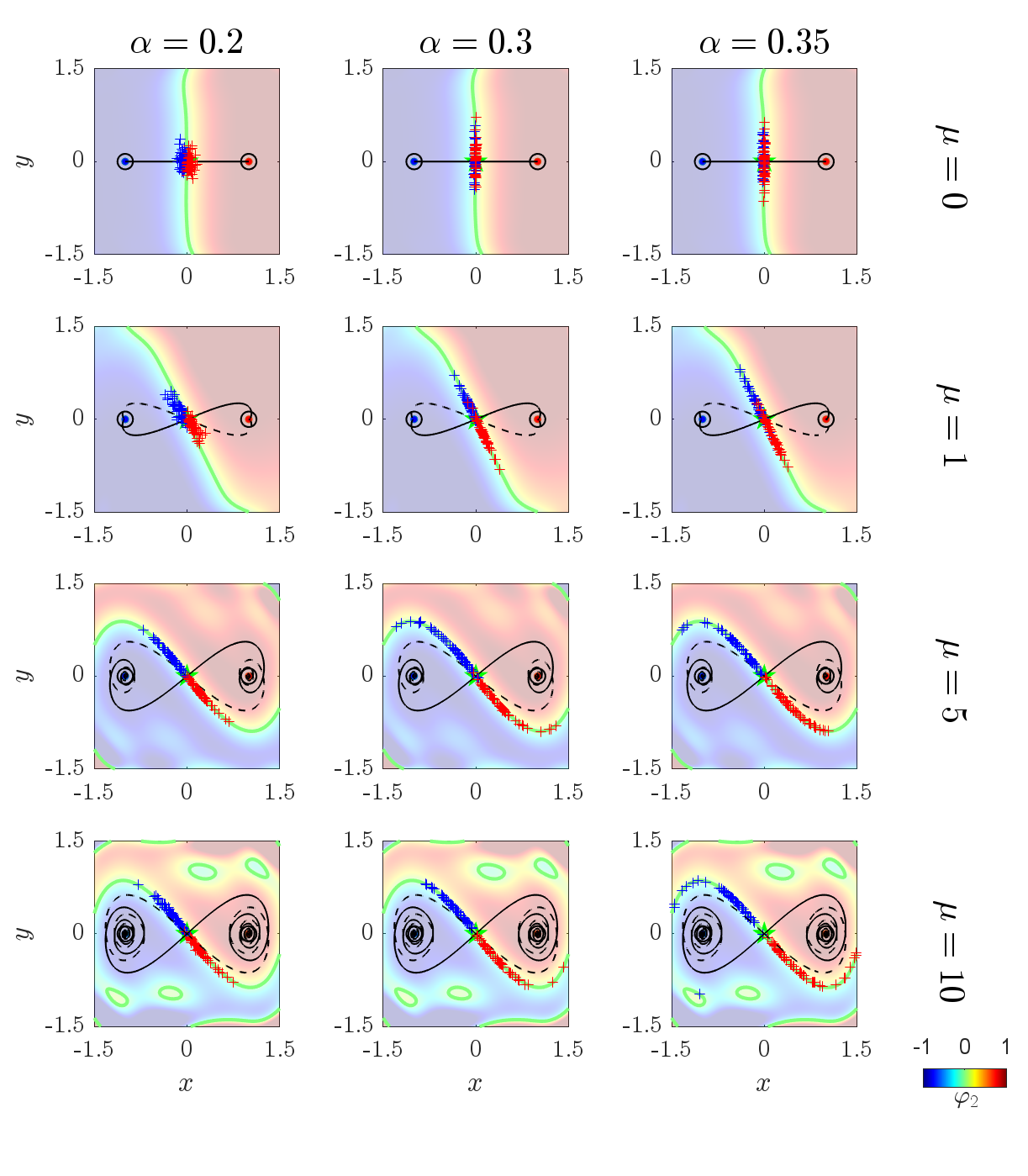}
    \caption{Crossing points (blue and red crosses) for different noise strengths $\alpha$ and rotation rate $\mu$ are plotted together with contours of $\varphi_2$. The blue circle, the red circle, and the green hollow star represent two attractors  and the saddle point, respectively. Red and blue colour shadings correspond to different level sets. The green line is the $0$-line, which defines the basin boundary.    
    } 
    \label{fig:all_mu_method_of_transition_2_}
\end{figure}

The presence of a good fit via large deviation theory of the relationship between mean exit times and noise intensity relies on the (implicit) assumption  that the trajectories cross the basin boundary near the saddle. To check whether this assumption holds in our case, we look into the spread of crossing points - identified as points in the space where $\varphi_2(\mathbf{x}(t))$ crosses zero - around the saddle for different strengths of noise and the rotation rates. We plot the detected crossing points together with the contours of $\varphi_2$ in Fig.~\ref{fig:all_mu_method_of_transition_2_}. Red and blue crosses indicate whether they were seeded around $(1,0)$ or $(-1,0)$, respectively. Light green contours represents basin boundaries for different $\mu$. Hollow light green markers at the origin represent the saddle. These plots shows that for no $(\mu=0)$ or weak rotation $(\mu=1)$, crossing invariably occurs in the close vicinity of the saddle even for a stronger noise intensity $(\alpha = 0.35)$. As we increase rotation $(\mu=5,~10)$, some crossings occur further away from the saddle. Since it is computationally expensive to reach the zero-noise limit, we choose noise intensities in such a way that, at least, most of the exit occurs through the region close to the saddle where the slope of the basin boundary is almost tangent to the saddle.

The spread of the crossing points for along the basin boundary $\mu=5$ and $10$  can be seen as a consequence of the fact that in the case of strong rotation the instantonic trajectories become almost tangent to the basin boundaries in a relatively vast region near the saddle. Hence, once a trajectory is closely aligned with an instantonic path, it can easily cross on to the other side of the basin boundary with a very moderate amount of noise. As we consider larger values of $\mu$, there is a growing separation in the magnitude of the drift in the direction orthogonal and parallel to $\nabla\Phi$. The avoidance of the saddle in noise-induced transitions for stochastic systems possessing multiscale deterministic component of the dynamics has been recently discussed in detail in \cite{borner_PRR:2024}. 

\begin{figure}[h]
    \includegraphics[width=0.75\linewidth, height=!]{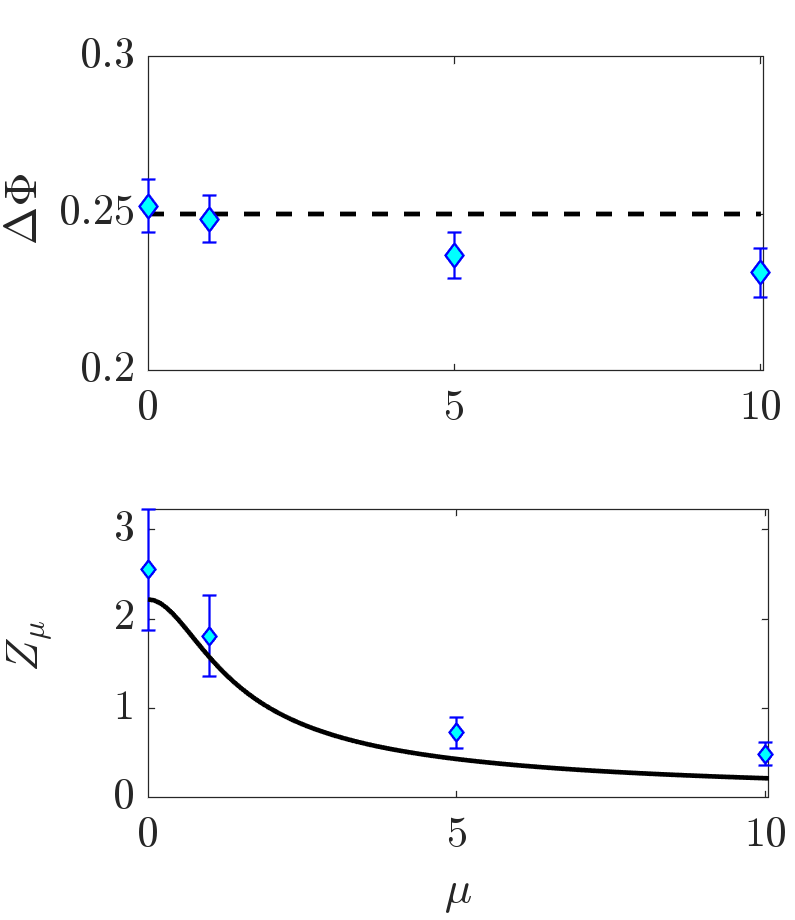}
    \caption{Height of the quasipotential, $\Delta \Phi$, (top) and the pre-exponential term,$Z_\mu$\ (bottom) are shown for different values of $\mu$. Errorbars indicate the $95\%$ confidence interval; see table~\ref{tab:coeff_stat_lin_fit}. The solid curves show the theoretical values of representation of $\Delta \Phi$ and $Z_\mu$.}
    \label{fig:transition_statistics_at_different_thresholds}
\end{figure}

Having a broader geometrical spread of the crossing points has an impact on our statistical analyses.  We plot the estimates of the $\Delta \Phi$ and $Z_\mu$ with the corresponding $95\%$ confidence interval in Fig.~\ref{fig:transition_statistics_at_different_thresholds} together with their theoretical values. The numerical values are reported in table~\ref{tab:coeff_stat_lin_fit}. The results show that the quality of the estimate of $\Delta \Phi$ and $Z_\mu$ deteriorates with increasing $\mu$ as crossings occur away from the saddle even for weaker noise intensities.

\begin{table}[h]
    \centering
    \begin{tabular}{|c|c|c|}
        \hline
        $\mu$ & $\Delta\Phi \pm E_{\Delta\Phi}$ & $Z_\mu \pm E_{Z_\mu}$ \\
        \hline
        \hline
        0 & 0.25253 $\pm$ 0.00422 & 2.55267 $\pm$ 0.33987\\
        \hline
        1 & 0.24834 $\pm$ 0.00375 & 1.80951 $\pm$ 0.22792\\
        \hline
        5 & 0.23667 $\pm$ 0.00366 & 0.72727 $\pm$ 0.08886\\
        \hline
        10 & 0.23126 $\pm$ 0.00395 & 0.49139 $\pm$ 0.06563\\
        \hline
    \end{tabular}
    \caption{Estimate of the height of the quasipotential and the prefactor with their $95\%$ confidence intervals. 
    }
    \label{tab:coeff_stat_lin_fit}
\end{table}

\begin{figure}[h]
    \includegraphics[width=0.75\linewidth, height=!]{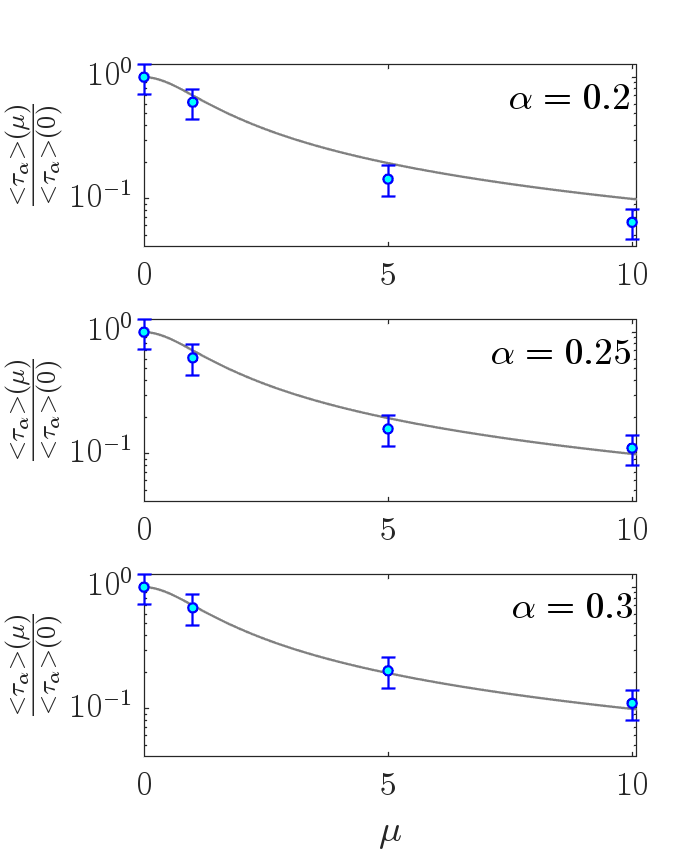}
    \caption{ Comparison between empirical (filled cyan circles) and analytical (grey curves) estimates of the ratio of mean exit times at any non-zero $\mu$ and $\mu=0$ for different noise intensities. This relationship can be associated with the dynamical properties of the saddle, see text. Errorbars in blue indicate the $95\%$ confidence interval. 
    }
    \label{fig:transition_statistics_of_ratio}
\end{figure}

By comparing Eq. \eqref{eqsaddle} with Eq. \eqref{eqn:ER_form_for_DWP}, we realise that it is possible to extract useful information by comparing directly $\langle \tau_\alpha \rangle(\mu)$ for different values of $\mu$ at a given value of $\alpha$. We have that $\langle \tau_\alpha \rangle(\mu)/\langle \tau_\alpha \rangle(0)=1/\sqrt{(1+\mu^2)}\approx\tau_{x_*}(\mu)/\tau_{x_*}(0)$ if $\alpha\ll1$. This provides an interesting link between the statistics of noise induced escapes and the life time of the saddle, which is an eminently deterministic quantity and does not depend on the intensity of the noise in the weak noise limit, as discussed before. This also clarifies that a key effect of increasing the absolute value of $\mu$ is the strengthening of the instabilities of the system. Positive confirmation of the relationship above is shown in Fig.~\ref{fig:transition_statistics_of_ratio} for different values of $\alpha$. 

\subsection{Role of the Saddle}
\label{subsec:Markov_model}
We have discussed above the importance of the saddle and its neighbouring region in mediating the transitions between the competing attractors and mentioned  that the trajectory lingers for a certain amount of time near the saddle - see Eq. \ref{eqn:modified_transition_time} - before a transition takes place. Yet, the latter effect is neglected in the large deviation theory-based analysis of noise induced transitions. In what follows we present an analysis that, instead, elucidates the statistical role of the saddle. 

Following the description given above on the mechanisms of transition between the competing metastable states, we approximate the system using a 3-state Markov chain, where state 1 corresponds to the neighbourhood of the equilibrium $(-1,0)$, state 2 corresponds to the neighbourhood of the equilibrium $(1,0)$, and state 3 corresponds to the neighbourhood of the saddle $(0,0)$. We assume that all transitions between state 1 and 2 have to go through state 3. At this level of coarse graining, one gets the following Master equation $\mathbf{\dot{p}}=\mathcal{M}\mathbf{p}$ for the populations $p_i$, $i=1,2,3$ of the three states:
\begin{align}
\dot{p_1}&=-kp_1+g p_3 \nonumber \\
\dot{p_2}&=-kp_2+g p_3 \nonumber\\
\dot{p_3}&=+kp_1+kp_2- 2 g p_3,
\label{eq:Markov_state_model}
\end{align} 
where $p_j$ is the population of the $j^{th}$ state (with $p_1p_2+p_3=1$), $k$ is the probability that a trajectory in the neighbourhood of a stable fixed point enters the neighbourhood of the saddle, whilst $g$ is the probability that a trajectory near the saddle relaxes towards either stable fixed points. In the formulation above, we respect the symmetry of the system described by Eq. \ref{eqn:Ito_diffusion_process}. Note that the Master equation given in Eq. \eqref{eq:Markov_state_model} does not allow for the existence of currents, so that it boils down to an equilibrium approximation of the system. From the discussion above, we consider the case where  $g\gg k$, because the system spends only a shorter time on the saddle than near the attracting fixed points. We obtain the following eigenpairs $\{\lambda_i,\mathbf{P}_i\}$, $i=1,2,3$ for the eigenvalue problem:
\begin{align}
\lambda_1=0&\quad \mathbf{P}_1=1/(2g+k)(g,g,k)^\top, \nonumber \\
\lambda_2=-k& \quad \mathbf{P}_2=(-1,1,0)^\top, \nonumber\\
\lambda_3=-2g-k& \quad \mathbf{P}_3=(-1,-1,2)^\top.
\end{align} 
$\mathbf{P}_1$ describes the invariant measure of the system, which is concentrated (symmetrically) on the states 1 and 2, with a much smaller projection on state 3. $\mathbf{P}_2$ describes the process of flipping between the state 1 and 2, whilst $\mathbf{P}_3$ describes the process of escape from the saddle towards the two attracting states.

Note that - see also discussion above - the mean lifetime in one metastable state before attempting to escape is given by $1/k$, whilst the mean time to actually transition to the other state is given by $2/k$. Instead, the lifetime of the saddle is approximately given by $1/(2g)$, as population is lost evenly to either state 1 and state 2 with rate $g$. 
See Eqs. \eqref{eqn:ER_form_for_DWP} and \eqref{eqsaddle} for an expression of $k$ and $g$ in terms of the full model.  
\begin{figure*}[t]
    \centering
    \includegraphics[width=.8\linewidth]{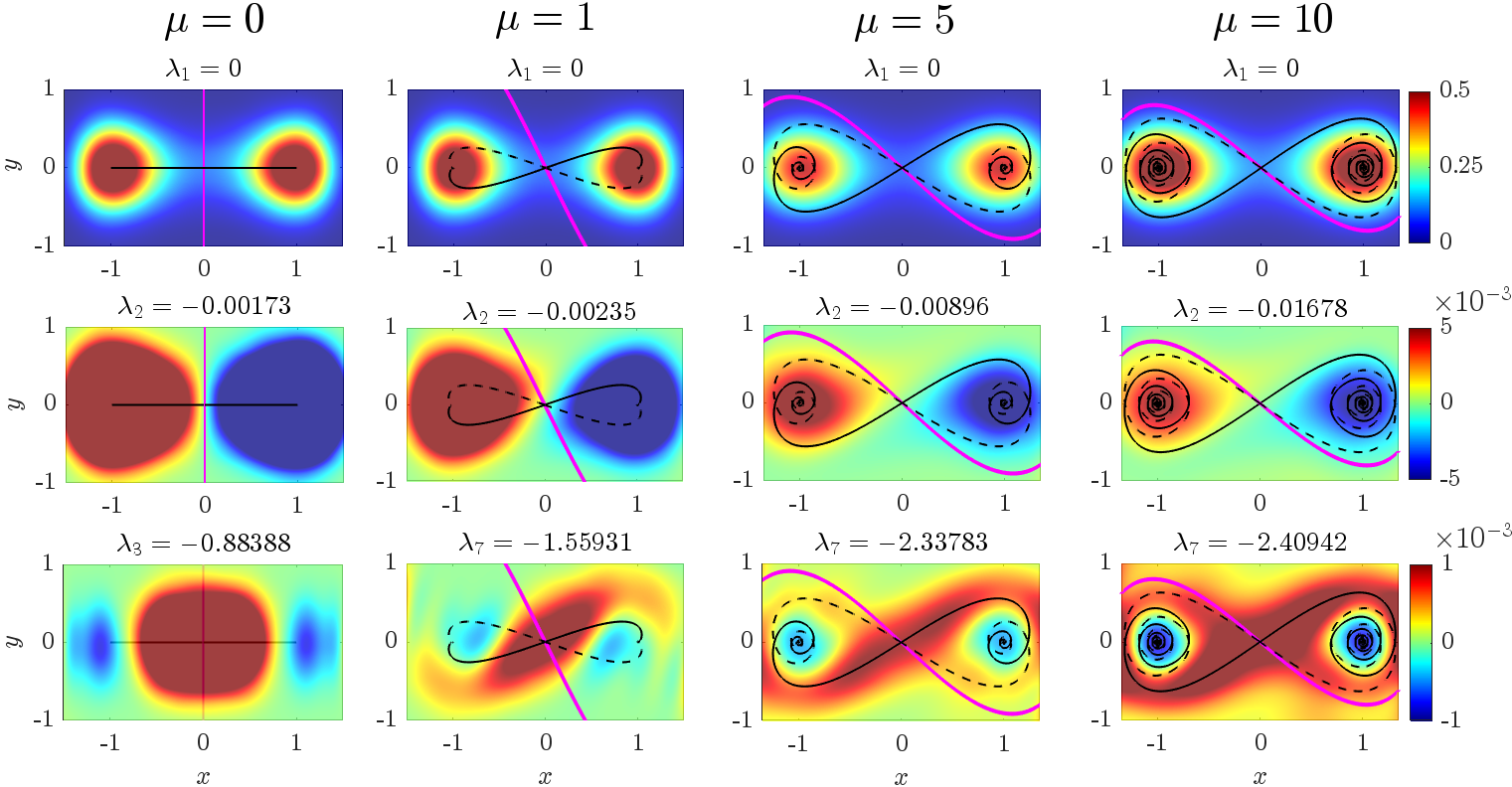}
    \caption{Eigenfunctions corresponding to the first three real eigenvalues of the Fokker-Planck operator corresponding to different strength of rotation, $\mu = 0,1,5,10$ (columns from left to right) and for $\alpha = 0.3$. Solid magenta lines represent the basin boundaries, dashed black curves represent the instantons connecting attractors and the saddle point, solid black lines describe relaxation trajectories from the saddle to the attractors.}
    \label{fig:PF_modes_structures}
\end{figure*}

It is possible to find the states corresponding to $\mathbf{P}_1$. $\mathbf{P}_2$, and $\mathbf{P}_3$ in the full model. Results are shown in Fig.~\ref{fig:PF_modes_structures} where we plot the first three (ordered in terms of the corresponding eigenvalue) real eigenmodes of the Fokker-Planck operator for $\alpha=0.3$ and for different values of $\mu$.

The first row of Fig.~\ref{fig:PF_modes_structures} shows the invariant measure $\rho_1(\mathbf{x})=\varphi_1^*(\mathbf{x})$ (corresponding to $\mathbf{P}_1$), which does not depend on $\mu$, in agreement with the fact that the Freidlin-Wentzell theory predicts that $\log(\rho_1(\mathbf{x}))\sim -2\Phi(\mathbf{x})/\alpha^2$, see discussion in \cite{margazoglou_ProcRSA:2021}. 

The second row of Fig.~\ref{fig:PF_modes_structures} shows the first subdominant mode $\varphi_2^*(\mathbf{x})$ (corresponding to $\mathbf{P}_2$), which is  associated with the transitions between the two metastable states (see the central column of Fig. \ref{fig:all_x2_comp} for the corresponding Koopman modes)  In this case the effect of rotation is apparent from the fact that $\lambda_2$ decreases with $\mu$, see Eq. \eqref{eqn:ER_form_for_DWP} and that the large-scale negative and positive density peaks are more concentrated near the stable fixed points, as the saddle becomes more repulsive. As $\mu$ becomes larger, the instantons and the relaxation trajectories become more and more distinct, see comment after Eq. \ref{eq:decomp_drift_vec}. The instantons tends to follow more closely the basin boundary, whilst the relaxation trajectories leave the saddle with an increasingly large angle with respect to the basin boundary. 

Finally, the third row of Fig.~\ref{fig:PF_modes_structures} shows the subdominant mode responsible for the escape of mass from the saddle towards the two competing metastable states (corresponding to $\mathbf{P}_3$). Note that the relaxation trajectories from the saddle (solid black lines) are approximately aligned with the positive mass anomaly of the mode centered on the saddle. Note also that this mode is the second subdominant mode ($\phi_3^*$) for the $\mu=0$ case, whilst it becomes the sixth subdominant mode ($\phi_7^*$) for the $\mu=1,$ 5, and 10 cases, because - see Table \ref{tab:coeff_stat_lin_fit_noise0p3} - the four rotational modes associated with the relaxation towards either attractor have a smaller real part. Processes associated with complex eigenvalues and eigenvectors cannot be represented with the 3-state simple Markov chain model, as mentioned above.

The existing literature has  stressed how looking at subdominant modes of the FP or Koopman operator allows one to identify quasi-invariant sets, where the system persists for a long time before transitions occur. Our results show that such a functional analytical framework also allows one to identify the much faster spontaneous relaxation process from saddle towards stable states, which, to our knowledge, had not been reported before, where the associated decay rate is related to the average time spent near the saddle.

\section{Summary and Conclusions}
\label{sec:conclusions}
In this work we have investigated the use of the Koopman formalism to study noise-induced transitions between competing metastable states. The motivation for this approach comes from the desire to be able to treat the problem in an equation- and geometry-agnostic way, with the ultimate goal of targeting high-dimensional complex system. 

The results obtained in this paper are just an initial step in this direction, as our study focuses on a prototypical symmetric bistable two-dimensional system driven by isotropic additive Gaussian noise of intensity $\alpha$. The drift term is given by the sum of minus the gradient of a double-well potential, plus an orthogonal component - proportional to $\mu$ times a rotation of the gradient of the double-well potential. The rotational component allows for the existence of macroscopic currents and brings the system out of equilibrium.

Overall, we present a methodology to analyze metastability in stochastic systems by identifying  slow inter-well and fast saddle escape processes in terms of Koopman modes. The approach avoids explicitly tracking the basin boundary, thus a geometry-agnostic framework, by computing subdominant Koopman mode from full-state data from a known two-dimensional model with very good sampling of both wells and the saddle. The subdominant Koopman mode provides a partition of the phase space into two metastable regions and is related to transitions between competing attractors. Such a mode can be used to characterize a) the boundaries between the deterministic basins of attraction; and b) 

The property a) indicates an alternative to  studying the committor function from the attractors. Specifically, the 0-level set of the submdominant mode the stable manifold of the saddle forming the basin boundary.

In this investigation, we have considered a symmetric bistable system. Yet, the criterion of the zero isoline of the first subdominant Koopman mode providing an accurate indication of the deterministic basin boundary applies also in the general asymmetric case and in arbitrary dimension. In this case, the basin boundary can be approximated with the isosurface of the committor function corresponding to a value that can be constructed from the relative occupancy of the two competing states, and is in general different from $1/2$ \cite{Mezic2020,Lohmann2025}. 

The strategy could in principle be generalised to systems possessing more than two competing states because in the weak noise limit one can identify the subdominant Koopman modes that govern the macroscopic behaviour of a system and are responsible for rare transitions between coexisting metastable states~\cite{froyland_NonlinDyn:2005,froyland_PRL:2007,lucarini_arXiv:2025}. If $N$  coexisting metastable states are present, one would need the time series of $N-1$ subdominant Koopman modes to identify the patwhays of transitions. We will pursue this line of work in future investigations.

The property b) comes from the fact that the time evolution of the subdominant Koopman mode constitutes a scalar observable that is able to identify a crossing of the basin boundary and consequently one can compute mean exit time statistics. This allows to bypass the geometrical complexity of attractors and saddles when studying transitions processes.

This allows us to estimate accurately the statistics of exit times from the attractors and find very good agreement with the prediction of large deviation theory in the weak noise limit, both in the equilibrium and in the nonequilibrium case. Note that agreement is found both for the subexponential prefactor and for the leading exponential factor.

We also provide strong numerical evidence of the recent theoretical result indicating that inverse of the absolute value of the subdominant Koopman eigenvalue is equal to the mean exit time from the attractor, both in the equilibrium and in the nonequilibrium case. This approach complements the estimation of mean exit times based on the time evolution of the subdominant Koopman mode.

Were we to consider an asymmetric bistable case, the link between the eigenvalue of the subdominant Koopman mode and the escape rates from either basin of attraction becomes less obvious. Yet, we can recover the desired agreement by changing the view point from a global to a local one. Indeed, if we consider the so-called killed process, whereby the stochastic process of interest is started in one basin of attraction and is forcibly stopped the moment it hits the basin boundary, its leading Koopman eigenvalue equals the inverse of the escape time \cite{le_CommMatPhys:2024}.  We will pursue this line of investigation in future works in order to better understand noise-induced tipping.

Further, investigating deeper Koopman modes we find that a set of subdominant eigenfunctions of the Koopman operators describes intra-well dynamics that can be associated with the linearized dynamics near the stable states. More interestingly, the Koopman formalism allows to define a mode that captures precisely the process of relaxation of the trajectories from the edge state towards the attractors. Such a process is implied but largely unaccounted for when using large deviation arguments. Moreover, the decay rate of this mode approximates the average time spent near the saddle, which has not been reported earlier, as per our knowledge. Uncovering this mode describing escape from the edge state towards attractors is another key finding of this work.

While our work has hopefully provided some novel ideas on how to study metastable systems by combining data-driven methods and functional analytical formalism, many questions still remain open. A pressing matter pertains finding a signature of the instantonic paths associated with transitions processes within Koopman modes. This could have a crucial importance for identifying precursors to tipping behavior and for better characterizing the resilience of a system. Another matter of great relevance is to extend the prototypical analysis presented here to higher dimensional systems, where we expect to beat the curse of dimensionality and the availability of limited data from system observables with the use of Koopman methods using Kernelized versions of the EDMD \cite{colbrook_chapter:2024}, by using the framework of residual DMD~\cite{colbrook_CommPAM:2024}, or by combining Koopman formalism with Markov state modelling \cite{lucarini_arXiv:2025}.

A final challenge comes from the fact that high-dimensional multistable system may well feature fractal basin boundary with co-dimension strictly smaller than one, resulting from the presence of complex, possibly multiscale dynamical processes \cite{lucarini_nonlinearity:2017,Bodai2020}. In this case, the zero-iso-surface of the Koopman mode (which is instead a regular manifold) provides a noise-regularized estimate of the basin boundary, and particular care should be placed in understanding the dynamics near the edge state.

\section*{Acknowledgement}
We wish to thank Dr. N. Zagli for valuable discussions on the draft. VL and MSG acknowledge the partial support provided by the Horizon Europe Project Past2Future (Grant No. 101184070). VL and AB acknowledge the partial support provided by the Horizon Europe Project ClimTIP (Grant No. 100018693). VL acknowledges the partial support provided by the ARIA SCOP-PR01-P003—Advancing Tipping Point Early Warning AdvanTip project, by the European Space Agency Project PREDICT (Contract 4000146344/24/I-LR), and by the NNSFC  International Collaboration Fund for Creative Research Teams (Grant No. W2541005).

\section*{Data Availability:} The data related to our findings are available in Figshare at \href{https://figshare.com/s/8fa4feca242d2ffe9ffc}{10.25392/leicester.data.32301609}.

\bibliography{bib_list}


\appendix

\section{EDMD algorithm for approximating the Koopman operator and its adjoint}
\label{sec:appendixI}
Here we recapitulate the key steps for the (finite) approximation of the family of Koopman operators $\{\mathcal{K}^t\}_{t \geq 0}$ following Williams \textit{et al.}~\cite{williams_JNSc:2015} and of the family of Perron-Frobenius operators (or, the Fokker-Planck semigroup) $\{\mathcal{P}^t\}_{t \geq 0}$ following Klus \textit{et al.}~\cite{klus_JCompDyn:2016}. Let us first assume that $\mathcal{F}$ be an infinite dimensional function space of square-integrable and complex-valued functions $f$ defined on the phase space $\mathcal{M}\subseteq \mathbb{R}^n$ \textit{s.t.} $\mathcal{F}=\{f|f:\mathcal{M} \rightarrow \mathbb{C}\}$. Operators $\mathcal{K}^t$ and $\mathcal{P}^t$ are defined on $\mathcal{F}$ s.t. $\mathcal{K}^t,\mathcal{P}^t:\mathcal{F} \mapsto \mathcal{F},~\forall t \geq 0$. Within the EDMD framework, we look for a finite-dimensional subspace of $\mathcal{F}$, called $\mathcal{F_D}$, to approximate $\mathcal{K}^t$ as a finite dimensional matrix $K$ and $\mathcal{P}^t$ as a finite dimensional matrix $P$. The set of finitely many functions spanning $\mathcal{F_D}$ are known as the dictionary set. Steps we follow to obtain $K$ and $P$ from timeseries data are as follows:
    \begin{enumerate}[(i)]
        \item For every pair of $(\mu,\alpha)$, we consider an ensemble of 100 simulated timeseries to sample a large number $(M)$ of snapshot pairs, large enough to be distributed with respect to the invariant measure. The total number of snapshot pairs $M$ is the sum of snapshot pairs $M_j,~j=1,2,...,100$ obtained from those 100 trajectories. For the $j^{th}$ ensemble member, every snapshot pair $(\pmb{x}_m,\pmb{x}_{m+1}) \textit{ s.t. } m=1,2,...,M_j$, are sampled from the timeseries with a sampling frequency $f_t$. Each element of these snapshot pairs are then grouped into two matrices, namely, $X^j_{M_j \times 2}=[\pmb{x}_1,\pmb{x}_2,...,\pmb{x}_{M_j}]^T$ and $Y^j_{M_j \times 2}=[\pmb{x}_2,\pmb{x}_3,...,\pmb{x}_{M_j+1}]^T$. The pair of matrices $(X^j_{M_j \times 2},Y^j_{M_j \times 2})$ has no relation with another pair $(X^i_{M_i \times 2},Y^i_{M_i \times 2})$ coming from a timeseries different from the $j^{th}$ one. Recall that, $\pmb{x}_m$ is the solution of the system Eq.~(\ref{eqn:Ito_diffusion_process}) at the $m^{th}$ time instance having an one-to-one correspondence with $\pmb{x}_{m+1}$ which is the solution at the next time instance. 
        
        By combining $X^j_{M_j \times 2},~(Y^j_{M_j \times 2})$,for $j=1,2,...,100$, we, thus, construct a snapshot matrix $X_{M \times 2},~(Y_{M \times 2})$ constituting $M$ elements. In the present analysis, the sample size (or, the total number of snapshot pairs, $M$) depends on the crossing time of the basin boundary for each trajectory. Therefore, the mean exit time $(\tau_\alpha \rangle(\mu))$ provides a lower bound for $M$: $M > \langle \tau_\alpha \rangle(\mu)/f_t *100$, where $f_t$ is the flight time and 100 samples used for each pair of $\mu,\alpha$. For choices of $\alpha \leq 0.3$ the sample length $(M)$ is in the order of $10^6-10^9$ depending on the values of $\mu$.
        \item A set of functions, forming a dictionary of basis functions, to approximate the Koopman operator is defined as ${\pmb{\Psi}}=[\psi_1, \psi_2, ... , \psi_{N_k}]$, where $\psi_k: \mathcal{M} \rightarrow \mathbb{C},~k=1,2,...,N_k$.
        \item By evaluating $X$ ($Y$) along dictionary functions we obtain a snapshot matrix $\Psi_x = {\pmb{\Psi}}(X)$ ($\Psi_y={\pmb{\Psi}}(Y)$) of the order $M \times N_k$. In this way, we uplift the data points from two-dimensional phase space to a higher-dimensional one. The number of snapshot pairs, $M$, depends on the sampling frequency which varies with respect to $\alpha$ and $\mu$. We choose $f_t$ in such a way that the computation of covariance matrices $G$ and $A$ is not too expensive, and at the same time convergence of the Galerkin approximation of the Koopman operator is achieved. Below  in Table~\ref{app_tab:samp_time_step_values} we list different values of $f_t$ used in the present analysis:
        \begin{table}[H]
        \centering
        \begin{tabular}{|c|c|c|c|}
            \hline
            \diagbox{$\mu$}{$\alpha$} & $0.2$ & $0.225$ & $\geq 0.25$ \\
            \hline
            0 & $2.5$ & $2$ & $0.005$\\
            \hline
            1 & $1.5$ & $1.5$ & $0.005$\\
            \hline
            5 & $1.5$ & $0.5$ & $0.005$\\
            \hline
            10 & $1.5$ & $0.01$ & $0.005$\\
            \hline
        \end{tabular}
        \caption Flight times $(f_s)$ used for computing EDMD matrices $K$ for different values of $\mu$ and $\alpha$.
        \label{app_tab:samp_time_step_values}
        \end{table}
        
        \item Now, to approximate $\mathcal{K}^t$ one need to find $K \in \mathbb{C}^{N_k \times N_k}$ which satisfies the relation $\Psi_y = \Psi_xK$. Therefore, the finite approximation of $\mathcal{K}^t$ over the space spanned by $\pmb{\Psi}$ is $K=\Psi_x^\dagger \Psi_y$, where $\dagger$ denotes the pseudo-inverse of a matrix. Since the pseudo inverse of a snapshot matrix is computationally expensive, two covariance matrices, namely $G$ and $A$, are generated such that $K=G^\dagger A$, where $G=\Psi_x^*\Psi_x$ and $A=\Psi_x^*\Psi_y$.
        \item The matrix $P \in \mathbb{C}^{N_k \times N_k}$ approximating the Perron-Frobenius operator $\mathcal{P}^t$ is obtained from the adjoint relationship between $\mathcal{K}^t$ and $\mathcal{P}^t$. Following this arguement, Klus \textit{et al.}~\cite{klus_JCompDyn:2016} showed that $P$ is similar to the Koopman matrix $K$ and $P=G^\dagger K^T G$ or, $P=G^\dagger A^{*}$. 
        \item Next we perform eigendecomposition of $K$ using the MATLAB routine `eig'~\cite{MATLAB_2024b} -- $(\sigma_j,\xi_j)~,j=1,2,...,N_k$ denote eigenvalue/left eigenvector pairs of $K$. Then Koopman eigenfunctions $\varphi_j$ are approximated as $\varphi_j(\pmb{x})=\pmb{\Psi}(\pmb{x})\xi_j$. Similarly, we compute eigenvalue/eigenvector pairs $(\bar{\sigma}_j,\xi^\star_j)~,j=1,2,...,N_k$ of $P$ and corresponding eigenfunctions of $\mathcal{P}_t$ are $\varphi^\star_j(\pmb{x})=\pmb{\Psi}(\pmb{x})\xi^\star_j$. 
    \end{enumerate}
\subsection{Numerical robustness of EDMD approximations:}
\begin{figure}[h]
    \includegraphics[width=\linewidth, height=!]{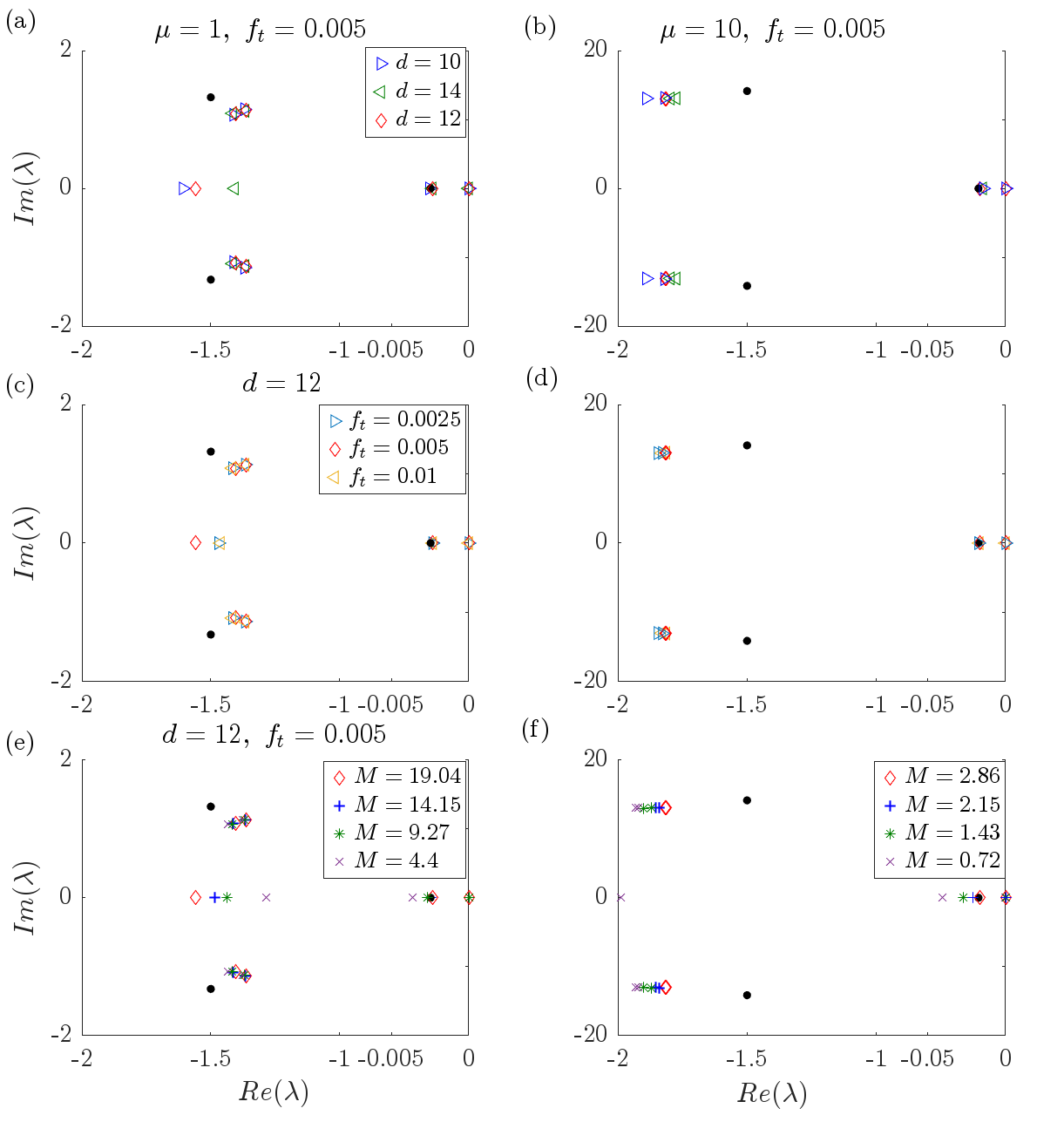}
    \caption{ Verification of the consistency of EDMD-eigenvalues for noise intensity $\alpha=0.3$, and for rotation $\mu=1$ (a, c, e) and $\mu=10$ (b, d, f) are shown. (a-b) Description of effects of the polynomial degrees as dictionary functions $d=$ 10 (blue triangles), 12 (red diamonds), and 14 (green triangles) for a fixed sampling time step $f_t=0.005$. (c-d) Effects of sampling time step are shown for three time steps $f_t=$ 0.0025 (blue triangles), 0.005 (red diamonds), and 0.01(yellow triangles) for $d=12$. (e-f) Effects of varying number of snapshot pairs are shown $M (\times 10^7)$ for four different values of M as represented by diamonds, cross, star, and diagonal cross markers. Solid black circles represent analytically obtained eigenvalues. 
    }
    \label{app_fig:eigenvalue_sensitivity}
\end{figure}

We also test the sensitivity of the first few eigenvalues of the Koopman matrix $K$ to EDMD parameters, namely, the highest degree of polynomial dictionary functions $(d)$, the sampling time step $(f_t)$, and the total number of snapshot pairs $(M)$. In this regard, we plot eigenvalues $(\lambda)$, satisfying $|Re(\lambda)|<2$, in Fig.~\ref{app_fig:eigenvalue_sensitivity} for different values of  $d,~f_t,$ and $M$. The left and right columns of Fig.~\ref{app_fig:eigenvalue_sensitivity} correspond to values $\mu=1$ and $10$, respectively, at $\alpha=0.3$. Since we know analytically six eigenvalues of $K$ ( $\lambda_1=0$, $\lambda_2$ from Eq.~\eqref{eqn:Matkowsky_identity}, $\lambda_{3,4,5,6}$ are conjugate pairs corresponding to eigenvalues of the linearized system Eq.~\eqref{eqn:Jacobian_RDWP} around $(-1,0)$ and $(1,0)$), we compare them with eigenvalues of approximated Koopman matrix $K$. In Fig.~\ref{app_fig:eigenvalue_sensitivity}, we only plot distinct analytical values (and drop the zero eigenvalue) indicated by solid black circle. First row, Fig.~\ref{app_fig:eigenvalue_sensitivity}(a-b), shows that eigenvalues are stable with respect to three different degrees of polynomial dictionary functions, $d=10$ (blue triangles), 12 (red diamonds) and 14 (green triangles), and are in close proximity to their analytical counterpart. Other EDMD parameters for these computations are kept fixed, $f_t=0.005$ and $M=19.04\times10^7$ for $\mu=1$, and $M=2.86\times 10^7$ for $\mu=10$, respectively. Second row, Fig.~\ref{app_fig:eigenvalue_sensitivity}(c-d), shows the stability of eigenvalues with respect to sampling time steps $f_t=0.0025$ (blue triangles), 0.005 (red diamonds) and 0.01 (yellow triangles) where we choose $d=12$. Again, eigenvalues closely align with analytical ones. Note that, $M$ also varies with changing $f_t$ but we keep the initial and final time for each ensembles fixed. Finally, Fig.~\ref{app_fig:eigenvalue_sensitivity}(e-f) show effects of varying number of samples on eigenvalues of $K$ computed for $f_t=0.005$ and $d=12$. Though $f_t$ is fixed, we change $M$ by varying the transient part of each trajectory. Larger $M$ means each trajectory has a smaller transient part to truncate whereas smaller $M$ means each trajectory has a larger transient part to be discarded - $+$, $*$, and $\times$ indicate transients as the initial ${1/4},~{1/2},$ and ${3/4}$ part of each trajectory, respectively. Keeping the sampling frequency fixed, the accuracy of eigenvalues deteriorates when we discard more than half of the data. This effects become more prominent for large $\mu$ as crossing is more frequent. 

\section{The Ulam approach}\label{sec:appendixII}
This section explains how to obtain the eigenvalues of the Koopman operator using Ulam's method. This algorithm has been extensively used in the computation of ergodic properties of dynamical systems \cite{froyland1999} and its application in statistical physics \cite{Lucarini2016,Tantet2018_jsp}. 

For each value of noise intensity $\alpha$ and rotation parameter $\mu$, the equations were integrated--- using an Euler-Maruyama method with $10^{-3}$ time step--- to obtain time series of $10^8$ snapshots separated by $5\times 10^{-2}$ time units. This integration time was sufficient for the system to sample both metastable states. The domain of phase-space $[-2,2]\times[-1.5,1.5]$ was discretized into $2^{12}$ equispaced boxes covering the entirety of the recorded snapshot data. Transitions between boxes were recorded into a Markov matrix $\mathbf{M}=(M_{ij})$ according to the following formula:
\begin{equation}\label{eq:transition_matrix}
    M_{ij} = \frac{\#\Big\{ \{\pmb{x}_k\}_{k=1}^{n-\tau} \in B_j  \land \{\pmb{x}_k\}_{k=\tau}^n\in B_i \Big\}}{\# \Big\{ \{\pmb{x}_k\}_{k=1}^{n-\tau} \in B_j\Big\}},
\end{equation}
where the positive integer $\tau$ is the flight time steps indicating the lag between snapshots before recording a transition and $B_i$ denotes a box for $i=1,\ldots,2^{12}$. The searching algorithm used here to locate the trajectory points in the box covering is based on a binary tree of the GAIO package \cite{gaio}. Since $\mathbf{M}$ is a Markov matrix, it represents the evolution of probability density functions on phase-space, hence, dual to the action of the Koopman operator. Nevertheless, it is expected that they both provide the same eigenvalues.

For noise values $\alpha>0.225$, a flight time of $\tau=2$ ($0.1$ time units) was enough to detect convergence in the eigenvalues. Convergence here means that the second largest-in-modulus eigenvalue is real, simple and with modulus strictly less than unity. For values $\alpha \leq 0.225$, a flight time of $\tau = 50$ steps was selected, to ensure that enough snapshot points experience transitions. The eignvalues of $\mathbf{M}$ depend on $\tau$, but their logarithm normalized by the flight time gives the estimates of the generator eigenvalues.

\end{document}